\documentclass[12pt,a4paper]{article}
\usepackage{amsmath,amsfonts,amsthm,graphicx}
\allowdisplaybreaks[3]
\numberwithin{equation}{section} 
\newtheorem{theorem}{Theorem}[section]

\newtheorem{definition}[theorem]{Definition}
\usepackage[body={15cm,21.5cm}]{geometry}
\newcommand{\cc}{{\mathbf c}}
\newcommand{\cd}{{\mathbf c}^{\dagger}}
\newcommand{\n}{{\mathbf n}}

\newcommand{\Qf}{{\mathbf Q}}
\newcommand{\Qb}{{\mathbb Q}}

\newcommand{\Tf}{{\mathbf T}}
\newcommand{\Tb}{{\mathbb T}}

\newcommand{\Lf}{{\mathbf L}}
\newcommand{\Lbf}{\overline{\mathbf L}}
\newcommand{\Lo}{\overline{L}}
\newcommand{\Rf}{{\mathbf R}}
\newcommand{\Ds}{{\mathsf D}}

\begin{document}
\title{
Asymptotic representations and 
$q$-oscillator solutions of the graded Yang-Baxter equation related to 
Baxter $Q$-operators
}
\author{Zengo Tsuboi
\footnote{
E-mail: ztsuboi$\bullet$yahoo.co.jp ; 
Present address: 
Department of Mathematics and Statistics, 
The University of Melbourne, Royal Parade, Parkville, Victoria 3010, Australia}
\\
Institut f\"{u}r Mathematik und Institut f\"{u}r Physik, \\
Humboldt-Universit\"{a}t zu Berlin, 
\\ 
Johann von Neumann-Haus, 
\\ 
Rudower Chaussee 25, 12489 Berlin, Germany
\\
} 
\date{}
\maketitle
\begin{abstract}
We consider a class of asymptotic representations of the Borel subalgebra of the  
quantum affine superalgebra  $U_{q}(\hat{gl}(M|N))$.  
This is characterized by 
Drinfeld rational fractions. 
In particular, we consider contractions of $U_{q}(gl(M|N))$ in the FRT formulation  
and obtain explicit solutions of the graded Yang-Baxter equation in terms of 
$q$-oscillator superalgebras. 
These solutions correspond to L-operators for Baxter Q-operators. 
We also discuss an extension of these representations to the ones for contracted algebras of $U_{q}(\hat{gl}(M|N))$ by 
considering the action of renormalized generators of the other side of the 
Borel subalgebra. 
We define model independent universal Q-operators as the supertrace of the universal R-matrix 
and write universal T-operators in terms of these Q-operators based on
 shift operators on the supercharacters. 
These include our previous work on $U_{q}(\hat{sl}(2|1))$ case \cite{BT08} in part, 
and also give a cue for the operator realization of our Wronskian-like formulas on T-and Q-functions 
in \cite{T09,Tsuboi:2011iz}. 
\end{abstract}
Journal Ref: Nucl. Phys. B 886 (2014) 1-30 \\
e-Print: arXiv:1205.1471 [math-ph]\\
DOI: 10.1016/j.nuclphysb.2014.06.017\\
Report number: HU-Mathematik-2012 - 06; HU-EP-12/13 
\section{Introduction} 
The Baxter Q-operators were introduced \cite{Bax72} by Baxter when he solved 
the 8-vertex model. Nowadays his method of Q-operators is 
 recognized as one of the most powerful tools in quantum 
integrable systems. 
In particular, Bazhanov, Lukyanov and Zamolodchikov \cite{BLZ97} defined Q-operators 
as the trace of the universal R-matrix 
over q-oscillator representations of the Borel subalgebra of the quantum 
affine algebra $U_{q}(\hat{sl}(2))$. Their work based on the q-oscillator algebra 
was generalized and developed for 
 various directions \cite{BHK02,Kulish:2005qc,Boos07,Kojima08,BT08,BGKNR10}. 

In our previous paper \cite{BT08}, we 
gave Q-operators for the quantum affine superalgebra $U_{q}(\hat{sl}(2|1))$. 
Our Q-operators in \cite{BT08} are universal in 
the sense that they do not depend on the quantum space 
 and can be applied for both lattice models and quantum field theoretical models as well. 
 We also proposed \cite{T09} an idea that there are $2^{M+N}$ kind of Baxter Q-functions for $U_{q}(\hat{gl}(M|N))$ case 
and gave Wronskian-like formulas on T-and Q-functions 
for finite \cite{T09} and infinite \cite{Tsuboi:2011iz} dimensional representations for any $(M,N)$
\footnote{We also proposed\cite{GKLT10} Wronskian-like formulas for infinite dimensional representations for $(M,N)=(4,4)$ case 
in the context of the $AdS_{5}/CFT_{4}$ spectral problem.}. 
The Q-function in \cite{T09} is labeled by the index set $I$, which is a subset of the set $\{1,2, \dots, M+N \}$. 
 In this paper, we continue these our previous works 
 and define model independent universal Q-operators for $U_{q}(\hat{gl}(M|N))$ (or  $U_{q}(\hat{sl}(M|N))$) as the 
supertrace of the universal R-matrix for any $(M,N)$. 
This gives a cue for the operator realization of the Wronskian-like formulas in \cite{T09,Tsuboi:2011iz}.

In section 2, we define the quantum affine superalgebra 
(or rather quantum loop superalgebra) $U_{q}(\hat{sl}(M|N))$ 
in terms of the Chevalley generators 
and the universal R-matrix associated with it. We also mention their extension to $U_{q}(\hat{gl}(M|N))$. 
Our task is basically evaluate this universal R-matrix for 
q-oscillator representations of the Borel subalgebra. 
As is well known, the Yang-Baxter equation follows from the 
 defining relations of the universal R-matrix. 
 The images of the universal R-matrix for particular representations give 
  the so-called L-operators and R-matrices. 
  The Yang-Baxter equations for the L-operators and the R-matrix ($RLL=LLR$ relations), 
which are also image of the Yang-Baxter equation for the universal R-matrix, 
  give another realization of the quantum affine superalgebra (FRT realization, \cite{Faddeev:1987ih}). 
  In accordance with the quantum affine superalgebra, 
 the quantum (finite) superalgebra $U_{q}(gl(M|N))$ also have these two realizations. 
In section 3, we consider $2^{M+N}$ kind of contractions  of the 
L-operator for $U_{q}(gl(M|N))$, which define contracted algebras  $U_{q}(gl(M|N; I))$. 
A preliminary form of these contractions for $(M,N)=(3,0)$ case was previously considered in 
\cite{Bp}. We also reported such contractions for $(M,N)=(2,1)$ case in conferences \cite{talks}. 

Next, we consider q-oscillator realizations of these 
contracted algebras. These induce representations of the Borel subalgebra of the quantum 
affine superalgebra (or q-superYangian) via the evaluation map. 
We remark that these representations can not be straightforwardly extended to the full quantum affine superalgebra.  
These are examples of  asymptotic representations characterized by the Drinfeld rational 
fractions
\footnote{They considered \cite{HJ11} an asymptotic algebra associated with the Drinfeld's second realization of 
the non-twisted quantum loop algebra.  
In this paper, we did not consider Ding-Frenkel type isomorphism from their algebra to our's, but rather 
developed our preliminary discussions \cite{talks,BT08,BT05} on L-operators for the Q-operators.} 
\cite{HJ11}. They are certain limits of the Kirillov-Reshetikhin modules (or their extension). 
The hart of an idea is to synchronize the highest weight of the representations and automorphisms of the 
algebra in the limit so that one can obtain finite quantities. 
In this way, we obtain spectral parameter dependent L-operators whose matrix elements are 
written in terms of the 
q-oscillator superalgebras.  Similar L-operators for 
$(M,N)=(3,0)$ were previously considered in \cite{BT05} and \cite{BGKNR10}. 
We also reported such L-operators for $(M,N)=(2,1)$ in \cite{talks,BT08}. 
All these L-operators satisfy the defining relations of the universal R-matrix (mentioned in section 2) 
 evaluated by the tensor product of the q-oscillator representations and the fundamental representation of the 
Borel subalgebras. It should be remarked here that the above q-oscillator representations of the Borel subalgebra can 
be extended to those of contracted algebras $U_{q}(\hat{gl}(M|N;I))$ of $U_{q}(\hat{gl}(M|N))$. For example for $U_{q}(\hat{gl}(2))$ case, 
the contracted algebra  $U_{q}(\hat{gl}(2;\{1 \}))$ in terms of the Chevalley generators is defined by 
the following commutation 
relations\footnote{In this paper, we need the level $0$ case of the algebra. One may extend this by adding 
the central element $c$ and the degree operator $d$: 
$h_{0}=k_{2}-k_{1}+c$, 
$[d,e_{i}]=\delta_{i,0}e_{i}$, $[d,f_{i}]=-\delta_{i,0}f_{i}$. 
The same remark can be applied for the higher rank case.}: 
\begin{align}
\begin{split}
& [e_{0}, f_{0}]=\frac{q^{h_{0}}}{q-q^{-1}}, 
\quad 
[e_{1}, f_{1}]=-\frac{q^{-h_{1}}}{q-q^{-1}}, 
\\
&[e_{0}, f_{1}]=[e_{1}, f_{0}]=[k_{i}, k_{j}]=0,
\\
&[k_{i}, e_{0}]=(\delta_{i,2}-\delta_{i,1})e_{0}, \quad 
[k_{i}, e_{1}]=(\delta_{i,1}-\delta_{i,2})e_{1},
\\
&[k_{i}, f_{0}]=-(\delta_{i,2}-\delta_{i,1})f_{0}, \quad 
[k_{i}, f_{1}]=-(\delta_{i,1}-\delta_{i,2})f_{1},
\\
& [e_{0}, [e_{0}, e_{1}]_{q^{-2 } } ]=
[e_{1}, [e_{1}, e_{0}]_{q^{2 } } ]=
[f_{0}, [f_{0}, f_{1}]_{q^{2} } ]=
[f_{1}, [f_{1}, f_{0}]_{q^{-2} } ]=0 ,
\end{split}
\end{align}
where $i,j \in \{1,2\}$, $h_{1}=-h_{0}:=k_{1}-k_{2}$, $[X,Y]_{q}:=XY-qYX$, $[X,Y]:=[X,Y]_{1}$. 
The generators of  the Borel subalgebra of $U_{q}(\hat{gl}(2;\{1\}))$ automatically satisfy 
the defining relations of the Borel subalgebras of $U_{q}(\hat{gl}(2))$. 
The restriction of the above relations to the generators $\{e_{1},f_{1},k_{1},k_{2}\}$ gives  $U_{q}(gl(2;\{1 \}))$. 
Then we can consider evaluation representations of $U_{q}(\hat{gl}(2;\{1\}))$ in terms of the representations of $U_{q}(gl(2;\{1 \}))$. 
The q-oscillator representations of the Borel subalgebra of $U_{q}(\hat{sl}(2))$ introduced by 
Bazhanov, Lukyanov and Zamolodchikov \cite{BLZ97}  are special cases of this type of representations. 

In section 4, we define the universal Q-operators as the supertrace of the universal R-matrix 
over the q-oscillator representations defined in the previous section. 
The T-operators are written in terms of these Q-operators. 
In the same way as previous paper \cite{BT08}, 
our Q-operators here are universal in the sense that they do not depend on the quantum 
space. 
As an example, we write Q-operators whose quantum space is the fundamental representation 
on each lattice site based on the L-operators derived in section 3. 
Section 5 is devoted to concluding remarks. 
Technical details are tucked into the appendices and a number of footnotes. 

There are many literatures on Q-operators related to $sl(2)$, which we could not refer. 
However there are not so many references for the higher rank case or  superalgebras case, 
which are our main subjects of this paper; and 
here we only mention some of them for rational models. 
In the rational limit ($q \to 1$; after multiplying diagonal matrices for the renormalization), our L-operators naturally reduce to L-operators which are similar to the ones
 proposed recently in \cite{BFLMS10} for rational lattice models.  
However, our L-operators are not simple generalization of the rational ones since 
many of the non-zero matrix elements of our L-operators become zero in the rational limit. 
Thus the q-deformation of the rational L-operators is not trivial. 
%
There are also Q-operators for infinite dimensional representations on the quantum
 space \cite{BDKM06,Frassek:2011aa}. 
 It will be interesting to evaluate our universal Q-operators for infinite dimensional representations on 
 the quantum space and to see how (or if) their formulas are lifted to the trigonometric case.  
We also proposed \cite{Kazakov:2010iu} Q-operators  based on the co-derivative \cite{KV07} on the supercharacters of 
$gl(M|N)$. This construction of the Q-operators is useful to discuss \cite{Kazakov:2010iu,AKLTZ11} 
functional relations among T-and Q-operators and embed them into the soliton theory. 
It is desirable to generalize this for the trigonometric case. 
\section{The quantum affine superalgebra 
$U_{q}(\hat{sl}(M|N))$ and the universal R-matrix, and their extension to $U_{q}(\hat{gl}(M|N))$}
\subsection{The quantum affine superalgebra $U_{q}(\hat{sl}(M|N))$ }
Let us introduce a grading parameter $p(i)=0$ for 
$i \in \{1,2,\dots, M\}$ and 
$p(i)=1$ for $i \in \{M+1,M+2,\dots, M+N \}$.
The quantum affine superalgebra $U_{q}(\hat{sl}(M|N))$ \cite{Yamane99} 
(see also \cite{KT94}) is a ${\mathbb Z}_{2}$-graded Hopf algebra 
generated by the generators
\footnote{In this paper, we do not use the degree operator $d$.}
 $e_{i},f_{i},h_{i}$, where 
$i \in \{0,1,\dots, M+N-1\}$. 
We assign the parity for these generators so that 
 $p(e_{0})=p(e_{M})=p(f_{0})=p(f_{M})= 1$ for $MN \ne 0$ and 
$p(X)=0$ for all the other generators $X$. 
For any $X,Y \in U_{q}(\hat{sl}(M|N))$, 
we define $p(XY)=p(X)+p(Y)$($\mathrm{mod}2$). 
We introduce the generalized commutator 
$[X,Y]_{q}=XY-(-1)^{p(X)p(Y)}q YX$. In particular, we set  
$[X,Y]_{1}=[X,Y]$.
For $i,j \in \{0,1,2,\dots,M +N-1 \}$, the
 defining relations of the algebra $U_{q}(\hat{sl}(M|N))$ are 
given by 
\begin{align}
& [h_{i},h_{j}]=0, \quad [h_{i}, e_{j} ] =a_{ij} e_{j}, \quad 
[h_{i}, f_{j} ] =-a_{ij} f_{j}, 
\\
&[e_{i},f_{j}]=\delta_{ij} \frac{q^{h_{i}} -q^{-h_{i}} }{q-q^{-1}}, 
\label{efk-1}
\\
& [e_{i},e_{j}]= [f_{i},f_{j}]=0 \quad \text{for} \quad a_{ij}=0, 
 \label{com-ee}
\end{align}
where $(a_{ij})_{0 \le i,j\le M+N-1}$ is the 
Cartan matrix 
\begin{align}
a_{ij}=((-1)^{p(i)}+(-1)^{p(i+1)})\delta_{ij}-
(-1)^{p(i+1)}\delta_{i,j-1}-(-1)^{p(i)}\delta_{i,j+1}, 
 \label{Cartan-mat}
\end{align}
here  $i,j$ should be interpreted modulo $M+N$: $p(M+N)=p(0)$, $\delta_{i,-1}= \delta_{i,M+N-1},
\delta_{i,M+N}= \delta_{i,0}$.
In addition to the above relations, there are Serre relations
\begin{align}
&[e_{i},[e_{i},e_{j}]_{q}]_{q^{-1}}=0, \quad 
[f_{i},[f_{i},f_{j}]_{q^{-1}}]_{q}=0  
\quad \text{for} \quad |a_{ij}|=1 , \quad 
a_{ii} \ne 0 , 
\label{Serre-aff}
\\
&[e_{i},[e_{i},[e_{i},e_{j}]_{q^{2}}]]_{q^{-2}}=0, \quad 
[f_{i},[f_{i},[f_{i},f_{j}]_{q^{-2}}]]_{q^{2}}=0  
\nonumber \\ 
& \qquad \text{for} \quad (M,N)=(2,0) , (0,2) , 
\quad i \ne j , 
\label{Serre-affgl2}
\end{align}
and also for the superalgebra case ($ MN \ne 0$), 
the extra Serre relations
\footnote{
\eqref{extraS} is equivalent to 
$[[e_{i},e_{j}]_{q} , [e_{k},e_{j}]_{q^{-1}}] = [[f_{i},f_{j}]_{q^{-1}} , [f_{k},f_{j}]_{q}] = 0 $ under $(e_{j})^{2}=(f_{j})^{2}=0$. 
We heard from Hiroyuki Yamane that 
we will need infinitely many Serre relations for $M=N=2$ case,  due to the Lusztig isomorphism 
(see \ \cite{Yamane99}, for more details).}:  
\begin{multline}
[\,[\,[e_{i},\, e_{j}]_{q },\, e_{k}]_{q^{-1}},\,e_{j}]=0 , \qquad 
[\,[\,[f_{i},\, f_{j}]_{q^{-1}},\, f_{k}]_{q},\,f_{j}]=0 
\\ \text{for}\quad 
  M+N \ge 4, \quad 
 (i,j,k)=(M+N-1,0,1),\quad (M-1,M,M+1), 
 \label{extraS}
\end{multline}
\begin{align}
[e_{0},\,[e_{2},\,[e_{0},\,[e_{2},\,e_{1}]_{q^{-1}}]]]_{q}&=
[e_{2},\,[e_{0},\,[e_{2},\,[e_{0},\,e_{1}]_{q^{-1}}]]]_{q},
\nonumber \\
[f_{0},\,[f_{2},\,[f_{0},\,[f_{2},\,f_{1}]_{q^{-1}}]]]_{q}&=
[f_{2},\,[f_{0},\,[f_{2},\,[f_{0},\,f_{1}]_{q^{-1}}]]]_{q}\ 
\quad \text{for} \quad (M,N)=(2,1) ,
\end{align}
\begin{align}
[e_{0},\,[e_{1},\,[e_{0},\,[e_{1},\,e_{2}]_{q^{-1}}]]]_{q}&=
[e_{1},\,[e_{0},\,[e_{1},\,[e_{0},\,e_{2}]_{q^{-1}}]]]_{q},
\nonumber \\
[f_{0},\,[f_{1},\,[f_{0},\,[f_{1},\,f_{2}]_{q^{-1}}]]]_{q}&=
[f_{1},\,[f_{0},\,[f_{1},\,[f_{0},\,f_{2}]_{q^{-1}}]]]_{q}\ 
\quad \text{for} \quad (M,N)=(1,2) , 
\label{extS3}
\end{align}
In this paper, we consider the case where the following
 central element is zero (level zero condition): 
\begin{align}
h_{0}+h_{1}+\cdots + h_{M+N-1}=0. 
\label{level0}
\end{align}
The algebra has the
 co-product $ \Delta : U_{q}(\hat{sl}(M|N)) \to U_{q}(\hat{sl}(M|N)) \otimes U_{q}(\hat{sl}(M|N))$ defined by 
\begin{align}
\Delta (e_{i})&=e_{i} \otimes 1 + q^{-h_{i}} \otimes e_{i}, \\
\Delta (f_{i})&=f_{i} \otimes q^{h_{i}} + 1 \otimes f_{i}, \\
\Delta (h_{i})&=h_{i} \otimes 1 + 1 \otimes h_{i}, 
\end{align}
where the tensor product is the graded one: 
$(A \otimes B)(C \otimes D)=(-1)^{p(B)p(C)}(AC \otimes BD)$. 
We assume that every tensor product $ \otimes $ in this paper is the graded one. 
We will also use an opposite co-product defined by
\begin{align}
\Delta'=\sigma\circ \Delta,\qquad \sigma\circ 
(X\otimes Y)=(-1)^{p(X)p(Y)}
Y\otimes X,\qquad X,Y\in U_{q}(\hat{sl}(M|N)).
\end{align}
In addition to these, there are anti-poide and co-unit, which will not be used in this paper. 

The Borel subalgebras ${\mathcal B}_{+}$ 
(resp. ${\mathcal B}_{-}$) is generated by 
$e_{i}, h_{i} $ (resp. $f_{i},h_{i}$), where 
$i \in \{0,1,\dots, M+N-1\}$. 
Let us take complex numbers $c_{i} \in {\mathbb C}$ which obey a relation
$\sum_{i=0}^{M+N-1} c_{i} =0$. Then the following transformation 
\begin{align}
h_{i} \mapsto h_{i} +c_{i} \quad \text{for} \quad 0 \le i \le M+N-1
 \label{shiftauto}
\end{align} 
gives a shift automorphism of $ {\mathcal B}_{+} $ or ${\mathcal B}_{-} $. 
Here we omit the unit element multiplied by the above complex numbers. 
This automorphism played a role
\footnote{When one takes a limit of the highest weight, one has to take a limit of these shift parameters 
at the same time to obtain a q-oscillator representation for the Q-operators.}
 in the construction of the Q-operators in \cite{BLZ97,BHK02,BT08}. 

There exists a unique element \cite{Dr85,KT92} 
${\mathcal R} \in {\mathcal B}_{+} \otimes {\mathcal B}_{-} $ 
called the universal R-matrix which satisfies the following 
relations
\begin{align}
\Delta'(a)\ {\mathcal R}&={\mathcal R}\ \Delta(a)
\qquad \text{for} \quad \forall\ a\in U_{q}(\hat{sl}(M|N))\,   ,\nonumber\\
(\Delta\otimes 1)\, 
{\mathcal R}&={\mathcal R}^{13}\, {\mathcal R}^{23}\, ,\label{R-def}\\
(1\otimes \Delta)\, {\mathcal R}&={\mathcal R}^{13}\, 
{\mathcal R}^{12}\,\nonumber 
\end{align}
where
\footnote{We will use similar notations for the L-operators 
to indicate the space where they non-trivially act on.} 
${\cal R}^{12}={\cal R}\otimes 1$, ${\cal R}^{23}=1\otimes {\cal R}$,
${\cal R}^{13}=(\sigma\otimes 1)\, {\cal R}^{23}$.
The (graded) Yang-Baxter equation 
\begin{align}
{\mathcal R}^{12}{\mathcal R}^{13}{\cal R}^{23}=
{\mathcal R}^{23}{\mathcal R}^{13}{\mathcal R}^{12}\ ,\label{YBE}
\end{align}
is a corollary of these relations \eqref{R-def}. 
The universal R-matrix can be written in the form
\begin{align}
{\mathcal R}=\overline{{\mathcal R}}\ q^{\mathcal K},
\qquad {\mathcal
  K}=\sum_{i,j=1}^{M+N-1}d_{ij}h_{i}\otimes h_{j},\label{R-red}
\end{align}
where $(d_{ij})_{1 \le i,j \le M+N-1}$ is the inverse 
of the Cartan matrix $(a_{ij})_{1 \le i,j \le M+N-1}$ 
of $sl(M|N)$. In case this Cartan matrix is degenerated ($M=N$), 
we have to consider an extended matrix
\footnote{
This may be achieved by adding an extra Cartan element 
$\sum_{j=1}^{M+N} (-1)^{p(j)} k_{j}$ to $U_{q}(\hat{sl}(M|N))$. Here $k_{j}$ are Cartan 
elements of $U_{q}(\hat{gl}(M|N))$, which we will introduce later. 
} and take the inverse of it 
\cite{KT91}. 
Here $\overline{{\mathcal R}}$  is the reduced universal $R$-matrix, which  
is a series in $e_j\otimes 1$ and $1 \otimes f_j$ 
and does not contain Cartan elements. 
Thus the reduced universal R-matrix is unchanged under the shift automorphism \eqref{shiftauto}, while 
the prefactor of the universal R-matrix \eqref{R-red} is shifted as 
\begin{align}
{\mathcal K}  \mapsto 
{\mathcal K}   + 
\sum_{i,j=1}^{M+N-1}d_{ij}c_{i} (1 \otimes h_{j}), 
\label{shiftpref}
\end{align}
where we considered a shift on ${\mathcal B}_{+} $.

\subsection{The quantum superalgebra $U_{q}(gl(M|N))$}
There is a (finite) quantum superalgebra $U_{q}(gl(M|N))$, which is generated 
by the elements $\{ e_{ij} \}_{i,j=1}^{M+N}$. We assign the parity 
of these generators as $ p(e_{ij})=p(i)+p(j) \mod 2$. 
Let us introduce the notations: 
$e_{\alpha_{i}}=e_{i,i+1}$,  $e_{-\alpha_{i}}=e_{i+1,i}$ 
for $i \in \{1,2,\dots, M+N-1 \}$. 
Then the defining relations of $U_{q}(gl(M|N))$ 
(for the distinguished simple root system) are (cf.\  \cite{KT91})
\footnote{
The last two relations 
$ [e_{ \pm \alpha_{M}},[e_{ \pm  \alpha_{M+1}}, [e_{ \pm  \alpha_{M}}, e_{ \pm \alpha_{M-1}} ]_{q^{\mp 1}}]_{q^{\pm 1}} ] =0$ 
 are equivalent to 
$[ [e_{ \pm  \alpha_{M}}, e_{ \pm  \alpha_{M-1}} ]_{q^{\mp 1}} , [e_{\alpha_{M}}, e_{\alpha_{M+1}} ]_{q^{\pm 1 }} ] =0$ 
under the condition $  (e_{\pm \alpha_{M}})^2=0 $.
} 
\begin{align}
& [e_{ii},e_{jj}]=0, \quad 
[e_{ii},e_{\pm \alpha_{j}}]=\pm (\delta_{i,j}-\delta_{i,j+1} ) e_{\pm \alpha_{j}}, 
\nonumber
\\ & 
[e_{\alpha_{i}}, e_{-\alpha_{j}}] = 
(-1)^{p(i)} \delta_{ij} 
 \frac{q^{ (-1)^{p(i)}e_{ii}- (-1)^{p(i+1)}e_{i+1,i+1}} - 
q^{- (-1)^{p(i)}e_{ii}+ (-1)^{p(i+1)}e_{i+1,i+1}}}{q-q^{-1}} , 
\nonumber
\\
& [e_{\alpha_{i}},e_{\alpha_{j}}]=
[e_{-\alpha_{i}},e_{-\alpha_{j}}]=0 \quad \text{for} 
\quad |i-j|\ge 2, 
\\
& [e_{\alpha_{i}},[e_{\alpha_{i}},e_{\alpha_{j}}]_{q}]_{q^{-1}}=
[e_{-\alpha_{i}},[e_{-\alpha_{i}},e_{-\alpha_{j}}]_{q^{-1}}]_{q}=0 
\quad \text{for}  \quad |i-j|= 1 \quad \text{and} \quad  p(e_{ \pm \alpha_{i}} ) =0 , 
\nonumber
\\ & (e_{\pm \alpha_{M}})^2=0  \quad \text{for}  \quad p(e_{ \pm \alpha_{M} }) =1,
\nonumber
\\
&  [e_{\alpha_{M}},[e_{\alpha_{M+1}}, [e_{\alpha_{M}}, e_{\alpha_{M-1}} ]_{q^{-1}}]_{q} ] =
[e_{-\alpha_{M}},[e_{-\alpha_{M+1}}, [e_{-\alpha_{M}}, e_{-\alpha_{M-1}} ]_{q}]_{q^{-1}} ] =0 
\nonumber 
\\
&
\qquad \text{for}  \quad   p(e_{ \pm \alpha_{M} }) =1 .
\nonumber 
\end{align}
The other elements are defined by 
\begin{align}
\begin{split}
e_{ij}&=[e_{ik},e_{kj}]_{q^{(-1)^{p(k)}}} \qquad 
\text{for} \quad i>k>j, \\
e_{ij}&=[e_{ik},e_{kj}]_{q^{-(-1)^{p(k)}}} \qquad 
\text{for} \quad i<k<j.
\end{split}
\end{align}
The other relations can also be obtain by \eqref{maprll1}-\eqref{maprll3} 
and \eqref{relationglmn1}-\eqref{relationglmn15}. 
Let $E_{ij}$ be a $(M+N) \times (M+N)$ matrix unit whose 
$(k,l)$-element is $\delta_{i,k} \delta_{j,l}$. 
The parity of this matrix is defined by 
$p(E_{ij})=p(i)+p(j) \mod 2$. 
$\pi(e_{ij})=E_{ij}$ gives the fundamental representation of 
$U_{q}(gl(M|N))$. 
There is an evaluation map
\footnote{$M=N=1$ case is special since \eqref{eva} does not satisfy \eqref{com-ee} 
for $(i,j)=(0,1),(1,0)$, in general.}
 $\mathsf{ev}_{x}$: 
$U_{q}(\hat{sl}(M|N)) \mapsto U_{q}(gl(M|N))$:  
\begin{align}
\begin{split}
& e_{0} \mapsto x q^{-(-1)^{p(1)}e_{11}} e_{M+N,1} q^{-(-1)^{p(M+N)}e_{M+N,M+N}},   \\
& f_{0} \mapsto (-1)^{p(M+N)}x^{-1} q^{(-1)^{p(M+N)}e_{M+N,M+N}} e_{1,M+N} q^{(-1)^{p(1)}e_{1,1}},  \\
& h_{0} \mapsto (-1)^{p(M+N)}e_{M+N,M+N}- (-1)^{p(1)}e_{1,1}, 
 \\
& e_{i} \mapsto e_{i,i+1}, \qquad 
f_{i} \mapsto (-1)^{p(i)}e_{i+1,i}, \qquad 
h_{i} \mapsto (-1)^{p(i)}e_{ii}- (-1)^{p(i+1)}e_{i+1,i+1}
  \\
& \qquad \text{for} \quad 1 \le i \le M+N-1,
\end{split} 
\label{eva}
\end{align}
where $x  \in {\mathbb C}$ is a spectral parameter. 
 
\subsection{Representations}
Let $\pi_{\lambda}$  be an irreducible representation 
of $U_{q}(gl(M|N))$ with the highest weight $\lambda =(\lambda_{1}, \lambda_{2}, \dots, \lambda_{M+N})$ 
and the highest weight vector $ |\lambda \rangle $ defined by 
\begin{align}
e_{ii} |\lambda \rangle =\lambda_{i} |\lambda \rangle,  \quad 
e_{jk} |\lambda \rangle =0 \quad \text{for} \quad  j<k, \quad 
\quad i,j ,k\in \{1,2,\dots, M+N \} .  
 \label{hwvglmn}
\end{align}
Then the composition $\pi_{\lambda }(x)=\pi_{\lambda } \circ \mathsf{ev}_{x}$ 
gives an evaluation representation of $U_{q}(\hat{sl}(M|N))$. 
For the fundamental representation, we will use a notation 
$\pi(x)=\pi_{(1,0,\dots,0) } (x)$.  
We also use a notation $\pi_{\lambda }^{+}(x)$ for the evaluation representation based on the 
Verma module defined by the free action of the generators on the highest weight 
vector \eqref{hwvglmn}. In this case, the representation is not necessary irreducible. 
Our main task is basically to evaluate the universal R-matrix for various representations of 
 $U_{q}(\hat{sl}(M|N))$ (or  $U_{q}(\hat{gl}(M|N))$). Namely, to find matrices of the form \eqref{R-red} which satisfy
 \eqref{R-def} for various representations of ${\mathcal B}_{+}$ and ${\mathcal B}_{-}$. 
The simplest example is the R-matrix for the Perk-Schultz model \cite{Perk:1981nb}  
 (see \cite{Cherednik80} for $N=0$ case), which is a multi-component generalization of the 
six-vertex model. 
Namely,  the image of the universal R-matrix 
for $\pi(x_{1}) \otimes \pi(x_{2}) $ gives (up to an overall factor $N(x_{1},x_{2})$; $x_{1},x_{2} \in {\mathbb C}$): 
\begin{align}
& \Rf (x_{1},x_{2})=N(x_{1},x_{2}) (\pi(x_{1}) \otimes \pi(x_{2})) {\mathcal R} = \Rf  -\frac{x_{1}}{x_{2}} \, \overline{\Rf}, 
 \label{PS-R}
\\
& \Rf = \sum_{i=1}^{M+N} q^{1-2p(i)} E_{ii} \otimes E_{ii} +
\sum_{i \ne j}  E_{ii} \otimes E_{jj} +
(q-q^{-1}) 
\sum_{i<j} (-1)^{p(j)} E_{ij} \otimes E_{ji},
\\
& \overline{\Rf} = \sum_{i=1}^{M+N} q^{-1+2p(i)} E_{ii} \otimes E_{ii} +
\sum_{i \ne j}  E_{ii} \otimes E_{jj} -
(q-q^{-1}) 
\sum_{i>j} (-1)^{p(j)} E_{ij} \otimes E_{ji} .
\end{align}
This obeys the graded Yang-Baxter equation
\begin{align}
\Rf^{12} (x_{1},x_{2})\Rf^{13} (x_{1},x_{3})\Rf^{23} (x_{2},x_{3})
=\Rf^{23} (x_{2},x_{3})\Rf^{13} (x_{1},x_{3})\Rf^{12} (x_{1},x_{2}),
\end{align}
which is an image of \eqref{YBE} for $\pi(x_{1}) \otimes \pi(x_{2})\otimes \pi(x_{3})$, where 
$x_{1}, x_{2}, x_{3} \in {\mathbb C}$. 

\subsection{Extension to $U_{q}(\hat{gl}(M|N))$}
Let us introduce Cartan elements $\{k_{i} \}_{i=1}^{M+N}$ of $U_{q}(\hat{gl}(M|N))$, which 
is related to the generators of $U_{q}(\hat{sl}(M|N))$ under \eqref{level0} as 
\begin{align}
\begin{split}
&h_{i}=(-1)^{p(i)} k_{i} - (-1)^{p(i+1)} k_{i+1}, \\
& [k_{i} ,k_{j}]=0, \qquad [k_{i},e_{j}] =(\delta_{ij} -\delta_{i,j+1} )e_{j}, 
\qquad [k_{i},f_{j}] =-(\delta_{ij} -\delta_{i,j+1} )f_{j}, 
\end{split}
\label{hatglmm}
\end{align}
where the indices $i,j$ should be interpreted modulo $M+N$. 
These are even elements $p(k_{i})=0$. 
It is sometimes convenient to define
\begin{align}
\overline{k}_{i}=-k_{i} \label{kbar} ,
\end{align}
and rewrite \eqref{efk-1}  as 
\begin{align}
[e_{i},f_{j}]=\delta_{ij} \frac{q^{(-1)^{p(i)}k_{i} +(-1)^{p(i+1)}\overline{k}_{i+1}} - q^{(-1)^{p(i)}\overline{k}_{i} +(-1)^{p(i+1)}k_{i+1}} }{q-q^{-1}} .
 \label{efk}
\end{align}
Later on, we will renormalize the generators and consider the case where 
$\overline{k}_{i}$ differs from $-k_{i} $ (cf.\ Appendix B). Moreover, this difference can be infinite in some limit. 
Now the Borel subalgebras $\mathcal{B}_{+}$ and $\mathcal{B}_{-}$ are generated by $\{e_{i},k_{i}\}$ and $\{f_{i},\overline{k}_{i}\}$, 
respectively.
The co-product is defined as $\Delta(k_{i}) =k_{i} \otimes 1 + 1 \otimes k_{i}$. 
For $M \ne N$, the pre-factor of the universal R-matrix \eqref{R-red} can be rewritten as 
\begin{align}
{\mathcal K}=\tilde{\mathcal K}-\frac{1}{M-N} {\mathcal C} \otimes {\mathcal C}, 
\end{align}
where ${\mathcal C} =\sum_{i=1}^{M+N} k_{i } $ is a central element and 
\begin{align}
\tilde{\mathcal K}=\sum_{i=1}^{M+N} (-1)^{p(i)} k_{i} \otimes k_{i}. 
\label{pre-R-re}
\end{align} 
Note that  
\begin{align}
\tilde{\mathcal R}= \overline{\mathcal R} q^{ \tilde{\mathcal K} }  
 \label{R-red-gl}
\end{align}
satisfies \eqref{R-def} for $U_{q}(\hat{gl}(M|N))$ generators. Then we 
regard\footnote{
To be precise, 
$\tilde{\mathcal R} q^{a {\mathcal C} \otimes {\mathcal C}}$ 
for any $a \in {\mathbb C}$ will be the universal R-matrix of $U_{q}(\hat{gl}(M|N))$ 
for \eqref{level0}. 
Here we normalized this for $a=0$. 
} 
 this renormalized universal R-matrix 
$\tilde{\mathcal R}$ 
as a universal R-matrix for $U_{q}(\hat{gl}(M|N))$ 
(under the condition \eqref{level0}). 
For $M \ne N$, $\tilde{\mathcal R} $ is related to $ {\mathcal R} $ via an overall central element: 
$\tilde{\mathcal R}= {\mathcal R} q^{\frac{1}{M-N} {\mathcal C} \otimes {\mathcal C} } $. 
However, $\tilde{\mathcal R}$ itself is 
 well-defined for $M=N$ case as well. 
 For any $c_{i} \in {\mathbb C}$ (multiplied by a unit element), 
the following transformation 
\begin{align}
k_{i} \mapsto k_{i} +(-1)^{p(i)} c_{i} \qquad \text{for} \quad 1 \le i \le M+N 
\label{shiftgl}
\end{align}
gives the shift automorphism of the Borel subalgebra. 
This keeps the level zero condition \eqref{level0} for any $c_{i} $. 
The prefactor of the universal R-matrix \eqref{R-red-gl} is shift by the shift automorphism \eqref{shiftgl} for $\mathcal{B}_{+}$ as 
\begin{align}
\tilde{\mathcal K} \mapsto \tilde{\mathcal K} +
\sum_{i=1}^{M+N} c_{i} (1 \otimes k_{i} ). 
\label{shift-pre-gl}
\end{align} 
The evaluation map for the Cartan elements is defined by 
\begin{align}
\mathsf{ev}_{x}(k_{i})=e_{ii} \qquad \text{for} \quad 1 \le i \le M+N. \label{eva-gl}
\end{align}
The evaluation representations are defined via this map in the same way as $U_{q}(\hat{sl}(M|N))$ case 
(the same symbols will be used). 
In particular $\pi(x)({\mathcal C})$ is a $(M+N) \times (M+N)$ unit matrix. 
In the subsequent sections, the contribution of 
the difference between  ${\mathcal R}$ and  $\tilde{\mathcal R}$ to each formula 
will be absorbed into a (representation dependent) overall factor of it. 
For example, the factor 
$(\pi(x_{1}) \otimes \pi (x_{2}) )(q^{\frac{1}{M-N} {\mathcal C} \otimes {\mathcal C} })= q^{\frac{1}{M-N}}$ for \eqref{PS-R} 
can be absorbed into $N(x_{1},x_{2})$. 
\section{L-operators from FRT realization of the quantum affine superalgebra $U_{q}(\hat{gl}(M|N))$}
\subsection{FRT realization of $U_{q}(\hat{gl}(M|N))$}
The quantum affine superalgebra $U_{q}(\hat{gl}(M|N))$ (and its subalgebra $U_{q}(gl(M|N))$)  
has another realization, called FRT realization \cite{Faddeev:1987ih} (see also, \cite{FM01}), 
based on the Yang-Baxter equation ($RLL=LLR$ relation). 
In this section we use this realization.
The (centerless) quantum affine  superalgebra  $U_{q}(\widehat{gl}(M|N))$ 
is defined by 
\begin{align}
& L_{ij}^{(0)}=\overline{L}_{ji}^{(0)}=0, 
\quad \text{for} \quad 1 \le i<j \le M+N 
 \label{rlla1} \\
& L_{ii}^{(0)} \overline{L}_{ii}^{(0)}=\overline{L}_{ii}^{(0)}L_{ii}^{(0)}=1
\quad \text{for} \quad 1 \le i \le  M+N,  
 \label{rlla2} \\
& \Rf^{23}(x,y)\Lf^{13}(y)\Lf^{12}(x)=
\Lf^{12}(x)\Lf^{13}(y)\Rf^{23}(x,y),
\label{rlla3}
\\
& 
\Rf^{23}(x,y)\Lbf^{13}(y)\Lbf^{12}(x)=
\Lbf^{12}(x)\Lbf^{13}(y)\Rf^{23}(x,y), 
\label{rlla4}
\\
& \Rf^{23}(x,y)\Lf^{13}(y)\Lbf^{12}(x)=
\Lbf^{12}(x)\Lf^{13}(y)\Rf^{23}(x,y),
\label{rlla5}
\end{align}
where $x,y \in {\mathbb C}$ and 
\begin{align}
\Lf(x)=\sum_{i,j=1}^{M+N} L_{ij}(x) \otimes E_{ij}, \quad 
\Lbf(x)=\sum_{i,j=1}^{M+N} \overline{L}_{ij}(x) \otimes E_{ij}, 
\end{align}
and 
\begin{align}
L_{ij}(x)=\sum_{n=0}^{\infty} L_{ij}^{(n)} x^{-n}, \quad 
\overline{L}_{ij}(x)=\sum_{n=0}^{\infty} \overline{L}_{ij}^{(n)}  x^{n}. 
\end{align}
The parity of the generators are defined by  $p(L^{(n)}_{ij})=
p(\overline{L}^{(n)}_{ij})=p(i)+p(j) \mod 2$.
The above relations came from the graded Yang-Baxter equation 
\eqref{YBE} for the universal R-matrix under the 
specialization \eqref{PS-R} and 
$\Lf (x)=N(x)(1 \otimes \pi(x)) \tilde{\mathcal R} $, 
$\Lbf (x)=\overline{N}(x)(1 \otimes \pi(x)) (\tilde{\mathcal R}^{21})^{-1} $, 
where $N(x)$ and $\overline{N}(x)$ are overall factors. 
In order to obtain the defining relations for  $U_{q}(\widehat{sl}(M|N))$, 
we will have to impose a condition that the quantum super-determinants of the above L-operators 
are 1.  But we do not impose this explicitly here. 
Let us introduce a function: $\theta(\text{True})=1$, 
$\theta(\text{False})=0$. 
One can rewrite \eqref{rlla3}  as 
\begin{multline}
(-1)^{(p(a)+p(b))p(c)}
(q^{(2p(a)-1)\delta_{ac}} x  - q^{(1-2p(a))\delta_{ac}} y)
L_{cd}(y) L_{ab}(x)
\\
-
(-1)^{(p(a)+p(b))p(d)}
(q^{(2p(b)-1)\delta_{bd}} x - q^{(1-2p(b))\delta_{bd}} y)
L_{ab}(x) L_{cd} (y)
= \\
= (-1)^{p(a) p(b)} (q-q^{-1})
\bigl[
\left( 
\theta(a>c) x +
\theta (a<c) y
\right) L_{ad}(y) L_{cb}(x) 
\\ 
-
\left( 
\theta(d>b) x +
\theta(d<b) y
\right) L_{ad}(x) L_{cb}(y) 
\bigr]
,
\end{multline}
and \eqref{rlla4}  as 
\begin{multline}
(-1)^{(p(a)+p(b))p(c)}
(q^{(2p(a)-1)\delta_{ac}} x  - q^{(1-2p(a))\delta_{ac}} y)
\Lo_{cd}(y) \Lo_{ab}(x)
\\
-
(-1)^{(p(a)+p(b))p(d)}
(q^{(2p(b)-1)\delta_{bd}} x - q^{(1-2p(b))\delta_{bd}} y)
\Lo_{ab}(x) \Lo_{cd} (y)
= \\
= (-1)^{p(a) p(b)} (q-q^{-1})
\bigl[
\left( 
\theta(a>c) x +
\theta (a<c) y
\right) \Lo_{ad}(y) \Lo_{cb}(x) 
\\ 
-
\left( 
\theta(d>b) x +
\theta(d<b) y
\right) \Lo_{ad}(x) \Lo_{cb}(y) 
\bigr]
,
\end{multline}
and \eqref{rlla5}  as 
\begin{multline}
(-1)^{(p(a)+p(b))p(c)}
(q^{(2p(a)-1)\delta_{ac}} x  - q^{(1-2p(a))\delta_{ac}} y)
L_{cd}(y) \Lo_{ab}(x)
\\
-
(-1)^{(p(a)+p(b))p(d)}
(q^{(2p(b)-1)\delta_{bd}} x - q^{(1-2p(b))\delta_{bd}} y)
\Lo_{ab}(x) L_{cd} (y)
= \\
= (-1)^{p(a) p(b)} (q-q^{-1})
\bigl[
\left( 
\theta(a>c) x +
\theta (a<c) y
\right) L_{ad}(y) \Lo_{cb}(x) 
\\ 
-
\left( 
\theta(d>b) x +
\theta(d<b) y
\right) \Lo_{ad}(x) L_{cb}(y) 
\bigr]
.
\end{multline}
For any $c \in \mathbb{C} \setminus \{0\}$, 
\begin{align}
\Lf(x) \mapsto \Lf(cx), \qquad 
\overline{\Lf}(x) \mapsto \overline{\Lf}(cx) 
 \label{shiftconst}
\end{align}
gives an automorphism of $U_{q}(\hat{gl}(M|N))$ since 
${\mathbf R}(cx_{1},cx_{2})={\mathbf R}(x_{1},x_{2})$. 
The restriction of the relations \eqref{rlla1}-\eqref{rlla5} to 
the relation for $\Lf(x)$ defines a sort of Borel subalgebra of 
$U_{q}(\hat{gl}(M|N))$  called $q$-super-Yangian. 
Note that the following transformation (multiplication of diagonal matrices in the second space) 
\begin{multline}
  \Lf(x) \mapsto  ( 1 \otimes \mathcal{H}_{L}) \Lf(x)  (1 \otimes \mathcal{H}_{R}), \quad 
 \overline{\Lf}(x) \mapsto  ( 1 \otimes \mathcal{H}_{L}) \overline{\Lf}(x)  (1 \otimes \mathcal{H}_{R}),
 \\
 \mathcal{H}_{L}=\sum_{i} \mathcal{H}_{L}^{(i)} E_{ii}, 
\quad \mathcal{H}_{R}=\sum_{i} \mathcal{H}_{R}^{(i)} E_{ii}, 
\quad  \mathcal{H}_{L}^{(i)} ,\mathcal{H}_{R}^{(i)}  \in {\mathbb C} \setminus \{ 0\}
 \label{shiftL} 
\end{multline}
keeps 
\footnote{
This is related to the parameters $c_{i}$ in the shift automorphism 
\eqref{shiftgl}-\eqref{shift-pre-gl} via $ \mathcal{H}_{R}^{(i)} =q^{c_{i}}$.
This also came from the first relation for the Cartan elements of $U_{q}(\hat{gl}(M|N))$ in \eqref{R-def}. 
If we restrict these Cartan elements to the ones for $U_{q}(\hat{sl}(M|N))$, 
we will obtain a restriction $\prod_{i=1}^{M+N} (\mathcal{H}_{L}^{(i)})^{(-1)^{p(i)}}=
\prod_{i=1}^{M+N} (\mathcal{H}_{R}^{(i)})^{(-1)^{p(i)}}=1$. 
In this case, \eqref{shiftL} (for $\mathcal{H}_{L}^{(i)} =1$) 
 should correspond to the shift automorphism \eqref{shiftauto} and 
\eqref{shiftpref}. 
Here we assumed that these parameters are not $0$ at first. 
However, 
we will have to consider limits that some of these go to $\infty$ or $0$.
} 
the relations \eqref{rlla1}  and \eqref{rlla3}-\eqref{rlla5}. 
However it  
changes \eqref{rlla2} as
\begin{align}
& L_{ii}^{(0)} \overline{L}_{ii}^{(0)}=\overline{L}_{ii}^{(0)}L_{ii}^{(0)}= (\mathcal{H}_{L}^{(i)}  \mathcal{H}_{R}^{(i)} )^{2}
\quad \text{for} \quad 1 \le i \le  M+N.
\end{align}
Then the inverse of $L^{(0)}_{ii}$ are not $\overline{L}_{ii}^{(0)}$ but renormalized 
generators $\overline{L}_{ii}^{(0)} (\mathcal{H}_{L}^{(i)}  \mathcal{H}_{R}^{(i)} )^{-2}$. 
We will meet a situation where some of $\overline{L}_{ii}^{(0)}$ diverge but $\overline{L}_{ii}^{(0)} (\mathcal{H}_{L}^{(i)}  \mathcal{H}_{R}^{(i)} )^{-2}$ remain finite in some limit. 
The restriction of this transformation to the q-superYangian
 gives an automorphism of it.   In addition, 
 if we consider a `bigger' algebra (a kind of an asymptotic algebra \cite{HJ11}) which does not assume \eqref{rlla2}, 
it can be an automorphism of such algebra.  

\subsection{FRT realization of $U_{q}(gl(M|N))$}
The quantum affine superalgebra  $U_{q}(\hat{gl}(M|N))$ has a finite subalgebra 
  $U_{q}(gl(M|N))$ defined by 
\begin{align}
& L_{ij}=\overline{L}_{ji}=0, 
\quad \text{for} \quad 1 \le i<j \le M+N 
\label{rll0}
\\
& L_{ii} \overline{L}_{ii}=\overline{L}_{ii}L_{ii}=1
\quad \text{for} \quad 1 \le i \le  M+N, 
\label{rll1} \\
& \Rf^{23}\Lf^{13}\Lf^{12}=\Lf^{12}\Lf^{13}\Rf^{23}, 
 \label{rll2}
\\
& \Rf^{23}\Lbf^{13}\Lbf^{12}=\Lbf^{12}\Lbf^{13}\Rf^{23},
\label{rll3}
\\ 
& \Rf^{23}\Lf^{13} \Lbf^{12}=\Lbf^{12} \Lf^{13}\Rf^{23}, 
\label{rll4}
\end{align}
where 
\begin{align}
\Lf=\sum_{i,j=1}^{M+N} L_{ij} \otimes E_{ij}, \quad 
\Lbf=\sum_{i,j=1}^{M+N} \overline{L}_{ij} \otimes E_{ij}. 
\label{rll4-2}
\end{align}
The parity of the generators are defined by  $p(L_{ij})=
p(\overline{L}_{ij})=p(i)+p(j) \mod 2$.
 Then the 
 relation \eqref{rll2} leads
\begin{multline}
(-1)^{(p(a)+p(b))p(c)}q^{(1-2p(a))\delta_{ac}} 
L_{cd} L_{ab}
-
(-1)^{(p(a)+p(b))p(d)}q^{(1-2p(b))\delta_{bd}} 
L_{ab}L_{cd}
= \\
= (-1)^{p(a) p(b)} (q-q^{-1})
\left( 
\theta(d<b) - \theta (a<c)
\right)
L_{ad} L_{cb}, \label{rll5}
\end{multline}
the relation \eqref{rll3} leads
\begin{multline}
(-1)^{(p(a)+p(b))p(c)}q^{(1-2p(a))\delta_{ac}} 
\Lo_{cd} \Lo_{ab}
-
(-1)^{(p(a)+p(b))p(d)}q^{(1-2p(b))\delta_{bd}} 
\Lo_{ab} \Lo_{cd}
= \\
= (-1)^{p(a) p(b)} (q-q^{-1})
\left( 
\theta(d<b) - \theta (a<c)
\right)
\Lo_{ad} \Lo_{cb}, \label{rll6}
\end{multline}
and the relation \eqref{rll4} leads
\begin{multline}
(-1)^{(p(a)+p(b))p(c)}q^{(1-2p(a))\delta_{ac}} 
L_{cd} \Lo_{ab}
-
(-1)^{(p(a)+p(b))p(d)}q^{(1-2p(b))\delta_{bd}} 
\Lo_{ab} L_{cd}
= \\
= (-1)^{p(a) p(b)} (q-q^{-1})
\left( 
\theta(d<b) \Lo_{ad} L_{cb} - 
\theta (a<c) L_{ad} \Lo_{cb}
\right)
. \label{rll7}
\end{multline}
For convenience, we list a more explicit form of these relations in Appendix A. 
These generators are related to the generators $\{ e_{ij} \}$ in section 2 as 
\begin{align}
& L_{ii}=q^{(-1)^{p(i)}e_{ii}}, 
 \label{maprll1}
\qquad 
\overline{L}_{ii}=q^{-(-1)^{p(i)}e_{ii}}, 
\\
& L_{ij}=(-1)^{p(i)}(q-q^{-1})e_{ji} q^{(-1)^{p(j)} e_{jj}} 
\quad \text{for} \quad i >j, \\
& \overline{L}_{ij}=-(-1)^{p(i)}(q-q^{-1})q^{-(-1)^{p(i)} e_{ii}}e_{ji}  
\quad \text{for} \quad i <j,  
\label{maprll3}
\end{align}

\subsection{Representations}
The action of generators of $U_{q}(gl(M|N))$ 
on the highest weight vector corresponding to \eqref{hwvglmn} is 
\begin{align}
\begin{split}
& L_{ii} |\lambda \rangle =q^{(-1)^{p(i)}\lambda_{i}} |\lambda \rangle,  \quad 
\overline{L}_{ii} |\lambda \rangle =q^{-(-1)^{p(i)}\lambda_{i}} |\lambda \rangle 
\quad \text{for} \quad 1 \le i \le M+N, 
\\
& L_{kj} |\lambda \rangle =0 \quad \text{for} \quad  1 \le j<k \le M+N .  
\end{split}
 \label{hwvglmn2}
\end{align}


There is an evaluation map from $U_{q}(\hat{gl}(M|N))$  
to $U_{q}(gl(M|N))$ such that
\begin{align}
\Lf (x)& \mapsto \Lf -\Lbf x^{-1}, \label{rll-ev1} \\
\Lbf (x)& \mapsto \Lbf -\Lf x.  \label{rll-ev2}
\end{align}
Apparently, the difference between $\Lf (x)$ and $\Lbf (x)$ 
 are not very important under the evaluation map. 
Let us consider an irreducible representation 
of $U_{q}(\hat{gl}(M|N))$ with the highest weight 
$(\nu(x),\overline{\nu}(x)) $ 
and the highest weight vector $ |\nu,\overline{\nu } \rangle $ defined by 
\begin{align}
&L_{ii}(x) |\nu,\overline{\nu }\rangle =\nu_{i}(x) |\nu,\overline{\nu }\rangle,  \quad 
\overline{L}_{ii}(x) |\nu,\overline{\nu } \rangle =\overline{\nu}_{i}(x) |\nu,\overline{\nu } \rangle 
\quad \text{for} \quad  1 \le i \le M+N, 
\label{weightl} \\
&L_{ij}(x) |\nu,\overline{\nu }\rangle =0,
 \quad 
\overline{L}_{ij}(x) | \nu,\overline{\nu } \rangle =0 \quad 
\text{for} \quad  i>j,
\end{align}
where 
$\nu(x)=(\nu_{1}(x),\nu_{2}(x), \dots, \nu_{M+N}(x))$, 
$\overline{\nu}(x)=(\overline{\nu}_{1}(x),\overline{\nu}_{2}(x), \dots, \overline{\nu}_{M+N}(x))$ 
are tuples of formal power series in $x^{-1}$ and $x$ respectively. 
For the evaluation representation based on \eqref{hwvglmn2}-\eqref{rll-ev2}, \eqref{weightl} becomes  
\begin{align}
&L_{ii}(x) |\lambda \rangle =(q^{(-1)^{p(i)} \lambda_{i}} - x^{-1} q^{-(-1)^{p(i)} \lambda_{i}} )|\lambda \rangle,  
\label{weightl20}
\\
& 
\overline{L}_{ii}(x) |\lambda \rangle =(q^{-(-1)^{p(i)} \lambda_{i} } - x q^{(-1)^{p(i)} \lambda_{i} } )|\lambda \rangle 
\quad \text{for} \quad  1 \le i \le M+N. 
\label{weightl2} 
\end{align}
For the finite dimensional representations, there exist monic polynomials in $x$, 
called Drinfeld polynomials
\footnote{Here we define these so that these become monic polynomials of 
the spectral parameter from ${\mathcal B}_{+}$. 
We can also define them so that they are monic polynomials of 
the spectral parameter from ${\mathcal B}_{-}$. In this case, $q$ in \eqref{Drinfeldp} will be replaced by $q^{-1}$. 
In addition, the definition for \eqref{defDrinfeld} for  $p(i)=1$ will have to be 
modified for the case where the Kac-Dynkin label 
take continuous number (typical representation). }
 $P_{i}(x)$, 
such that 
\begin{align} 
\frac{\nu_{i}(x^{-1})}{\nu_{i+1}(x^{-1})} =q^{(-1)^{p(i)} \mathrm{deg}P_{i}(x) }
 \frac{P_{i}(xq^{-2(-1)^{p(i)}}) }{P_{i}(x)} 
 =\frac{\overline{\nu}_{i}(x^{-1})}{\overline{\nu}_{i+1}(x^{-1})} 
 \qquad \text{for} \quad 1 \le i \le M+N-1.
 \label{defDrinfeld}
\end{align}
For the evaluation modules whose highest weights are  given by \eqref{weightl20} and \eqref{weightl2}, 
the Drinfeld polynomials have the form (if $\lambda_{i}-(-1)^{p(i)+p(i+1)} \lambda_{i+1} \in {\mathbb Z}_{\ge 0} $)
\begin{multline} 
P_{i}(x) = \prod_{k=1}^{\lambda_{i}-(-1)^{p(i)+p(i+1)} \lambda_{i+1}} 
(1-xq^{-2(-1)^{p(i+1)} \lambda_{i+1} -2(-1)^{p(i)} (k-1) }) 
\\
\text{for} \quad 1 \le i \le M+N-1. 
\label{Drinfeldp}
\end{multline}
For $N=0$ case, finite dimensional modules which are characterized by the Drinfeld polynomials with the condition 
$\lambda_{i}-\lambda_{i+1} =m \delta_{ik}$ (for all $1 \le i \le M-1$, and some $m \in \mathbb{ Z}_{\ge 0}$ and 
$1 \le k \le M-1$) 
are called Kirillov-Reshetikhin modules. 
\subsection{Contraction of $U_{q}(gl(M|N))$}
Let us take a subset $I$ of the set $\{1,2,\dots, M+N \}$ and 
 its complement set 
$\overline{I}:=\{1,2,\dots, M+N \} \setminus I$. 
There are $2^{M+N}$ choices of the subsets in this case. 
Corresponding to the set $I$, we consider $2^{M+N}$ kind of representations of the q-superYangian. 
For this purpose, 
we consider $2^{M+N}$ kind of  contractions of  $U_{q}(gl(M|N))$. 
At first, we change the condition \eqref{rll1} and define a contracted algebra as follows. 
\begin{definition}
The contracted algebra 
$\tilde{U}_{q}(gl(M|N;I))$ is an associative algebra over ${\mathbb C}$ with 
a unit element $1$ and generators $L_{ij}, \Lo_{ij}$ obeying 
 the relations 
\eqref{rll0}, \eqref{rll2}-\eqref{rll4-2} and 
\begin{align}
& L_{ii} \Lo_{ii}=\Lo_{ii}L_{ii}=1
\quad \text{for} \quad i \in I, 
 \label{red1} 
\\
& \Lo_{ii}=0 \qquad \text{for} \quad i \in \overline{I} . 
\label{red2} 
\end{align}
In addition, we assume the existence of an inverse element $L^{-1}_{ii} $ of $L_{ii}$ for any $i \in \{1,2,\dots, M+N\}$.
\begin{align}
L_{ii} L^{-1}_{ii} =L^{-1}_{ii} L_{ii} =1.
\end{align}
\end{definition}
Note that $L^{-1}_{ii} $ coincides with $\Lo_{ii}$ only for $i \in I$. 
Then one can obtain $2^{M+N}$ kind of algebraic solutions of the graded 
Yang-Baxter equation via the map \eqref{rll-ev1}. 
In addition to the contraction \eqref{red2}, 
we consider the following subsidiary contraction and define a contracted algebra which 
is smaller than $\tilde{U}_{q}(gl(M|N;I))$. 
\begin{definition}
Suppose the set $I$ has the form 
$I=\{k+1,k+2,\dots, k+n \}$ for some $k \ge 0, n >0$, then the contracted algebra $U_{q}(gl(M|N;I))$ 
is defined by adding the following relations to $\tilde{U}_{q}(gl(M|N;I))$. 
\begin{align}
L_{ij} &= 0 \qquad \text{for} 
 \quad k+n < i \le M+N  \quad \text{and} \quad  1 \le  j \le k, 
\label{red30} \\
\Lo_{ij} &= 0 \qquad \text{for} \quad 1<i<j \le k \quad \text{or} \quad k+n < i<j \le M+N .  \label{red3}
\end{align}
\end{definition}
One may consider different contractions than \eqref{red30}, \eqref{red3}. 
Here we consider a contraction so that the location of the zeros  
becomes cyclic with respect to the shift of the suffixes by 
an operation:
$a \mapsto a+1$ for $a<M+N$ and $M+N \mapsto 1$. Namely, the contraction for 
$k>0$ can be given by applying this operation $k$-times for the case $k=0$. 
What is important here is to respect the relations among the generators 
\eqref{rll5}-\eqref{rll7}. 
Let us apply the contraction \eqref{red2} to the relation \eqref{relationglmn14} 
for the case $a,b \in \overline{I}$. 
Then we obtain $[L_{ba}, \Lo_{ab}]=0$. This relation holds true automatically if $L_{ba}=0$ or $\Lo_{ab}=0$. 
This is an origin of our subsidiary contractions \eqref{red30}-\eqref{red3}. 
Thus the contractions (corresponding to \eqref{red30}-\eqref{red3}) for a generic set would be 
realized by putting one of  $ L_{ab}$ and $ \Lo_{ab}$ to $0$ for $a,b \in \overline{I} $.
Whether the contracted algebras for the generic sets
 have non-trivial useful representations is an open problem. 
For the contracted algebra $\tilde{U}_{q}(gl(M|N;I))$,  
the conditions \eqref{red30}-\eqref{red3} may hold true only on the level of representation. 
We remark that these contractions  on the L-operator for 
$U_{q}(\hat{gl}(3))$ (written in terms of the generators 
$e_{ij}$ and substituted into \eqref{rll-ev1}) 
was previously considered in \cite{Bp}. 
We also reported these contractions for $U_{q}(\hat{gl}(2|1))$ 
 in conferences \cite{talks}. 

\subsection{Representations of the contracted algebras}
The next task is to consider representations of these 
contracted algebras. 
We are interested in q-oscillator representations. 
The $q$-oscillator (super)algebra (see for example, \cite{Chaichian:1989rq}) is generated by the 
generators $\cc_{ai},\cd_{ia}, \n_{ia}$ for $i \in I$, 
$a \in \overline{I}$,  
whose parities are defined by 
$p(\cc_{ai})=p(\cd_{ia})=p(a)+p(i) \mod 2$, $p(\n_{ia})=0$. 
They obey the following defining relations: 
\begin{align}
& [\cc_{ai}, \cd_{jb}]_{q^{(-1)^{p(a)}\delta_{ab}\delta_{ij}}}  =\delta_{ab} \delta_{ij} q^{- (-1)^{p(i)} \n_{ia}}, 
\quad 
 [\cc_{ai}, \cd_{jb}]_{q^{-(-1)^{p(a)}\delta_{ab}\delta_{ij}}}  =\delta_{ab} \delta_{ij} q^{(-1)^{p(i)} \n_{ia}} , 
\label{qosc}
\\
&  
[\n_{ia}, \cc_{bj}]=-\delta_{ij}\delta_{ab} \cc_{bj}, \quad 
 [\n_{ia}, \cd_{jb}]=\delta_{ij}\delta_{ab} \cd_{jb}, \quad
[\n_{ia}, \n_{jb}]=[\cc_{ai}, \cc_{bj}]=[\cd_{ia}, \cd_{jb}]=0,
\end{align}
where  $i,j \in I$, $a,b \in \overline{I}$. 
From  \eqref{qosc}, 
we can derive the relations:  
 $ \cc_{ai}\cd_{ia}=[\n_{ia}+1]_{q}$, 
$ \cd_{ia}\cc_{ai}=[\n_{ia}]_{q}$ 
for $p(i)+p(a)=0 \mod 2$, and 
$ \cc_{ai}\cd_{ia}=[1-\n_{ia}]_{q}$,
$\cd_{ia}\cc_{ai}=[\n_{ia}]_{q}$  
for
\footnote{We consider these generators on 
the Fock space fixed by the vacuum \eqref{vacosc}. 
Then for the fermionic case $p(i)+p(a)=1 \mod 2$, 
these relation effectively becomes 
$ \cc_{ai}\cd_{ia}=1-\n_{ia}$,
$\cd_{ia}\cc_{ai}=\n_{ia}$. 
} $p(i)+p(a)=1 \mod 2$,
where $[x]_{q}=(q^{x}-q^{-x})/(q-q^{-1})$. 
Note that the following transformation 
\begin{align}
& 
\n_{ia} \mapsto \n_{ia}, 
\quad 
\cc_{ai} \mapsto \xi_{ia} \cc_{ai} q^{\sum_{(j,b) \in I \times \overline{I}} \eta^{jb}_{ia} \n_{jb}}, 
\quad 
\cd_{ia} \mapsto \xi_{ia}^{-1} q^{-\sum_{(j,b) \in I \times \overline{I}} \eta^{j,b}_{ia} \n_{jb}} \cd_{ia} , 
\nonumber 
\\
& \eta^{jb}_{ia}=\eta^{ia}_{jb} \in {\mathbb C}, 
\quad \xi_{ia} \in {\mathbb C} \setminus \{0\}, 
\quad 
i,j \in I, \quad  a,b \in \overline{I} 
 \label{autoosc}
\end{align}
gives a $|I||\overline{I}| (|I||\overline{I}| +3)/2 $ parameter continuous automorphism of 
the q-oscillator algebra \eqref{qosc}. We also remark that 
the following transformation 
\begin{multline}
\n_{ia} \mapsto -\n_{ia}-(-1)^{p(i)+p(a)} ,  
\qquad 
\cc_{ai} \mapsto  \cd_{ia}, 
\qquad 
\cd_{ia} \mapsto -(-1)^{p(i)+p(a)} \cc_{ai} 
%
\label{autoosc2}
\end{multline}
gives a discrete automorphism of the q-oscillator algebra \eqref{qosc} 
for any $i \in I $ and $ a \in \overline{I}$.
For the diagonal part, we consider the following
\footnote{For ${\mathbf L}$, this satisfies a $U_{q}(sl(M|N))$-type relation 
$\prod_{i \in I } L_{ii}^{(-1)^{p(i)}}  \prod_{a \in \overline{I} } L_{aa}^{(-1)^{p(a)}} =1$, 
but for ${\mathbf \Lo}$, it does not. } 
\begin{align}
L_{ii}&=q^{-(-1)^{p(i)}\sum_{b \in \overline{I}} \n_{ib}} \quad \text{for} \quad i \in I, 
\label{dia1} \\
L_{aa}&=q^{(-1)^{p(a)} \sum_{j \in I} \n_{ja}} \quad \text{for} \quad a \in \overline{I}, \\
\Lo_{ii}&=q^{(-1)^{p(i)} 
\sum_{b \in \overline{I}} \n_{ib}} \quad \text{for} \quad i \in I.
\label{dia3}
\end{align}
Let us look for q-oscillator realization of the 
non-diagonal part, which are compatible with the defining relations with 
 the diagonal part \eqref{dia1}-\eqref{dia3}. 
Let us introduce notations 
$\n_{[i,j],a}=\sum_{k=i}^{j} \n_{k,a} $, 
$\n_{i,[a,b]}=\sum_{c=a}^{b} \n_{i,c} $, 
$\n_{I,a}=\sum_{k \in I} \n_{k,a} $, 
$\n_{i,\overline{I}}=\sum_{c \in \overline{I}} \n_{i,c} $. 
We find the following solutions
\footnote{We used relations in Appendix A for the direct calculations.}. 

{\em (i) The case} 
$I=\emptyset$, $\overline{I}=\{1,2,\dots, M+N\}$: 
for $a,b \in \overline{I}$, 
\begin{align}
& L_{ab}=0 \quad \text{for} \quad a \ne b \quad 
\text{and} \quad 
L_{aa}=1, \\[5pt]
& \Lo_{ab}=0 . 
\end{align}

{\em (ii) The case}  
$I=\{ i \}$, $\overline{I}=\{1,2,\dots, M+N\} \setminus \{ i \}$:  
\begin{align}
& L_{\alpha \beta}=0 \quad \text{for} 
\quad \alpha < \beta \quad 
\text{or} \quad 1 \le \beta < i < \alpha \le M+N,  \\
& L_{ii}= q^{-(-1)^{p(i)} \n_{i,\overline{I}}}, 
\\
&L_{aa}=q^{(-1)^{p(a)}  \n_{i,a}} 
\qquad \text{for} \quad a \in \overline{I}, \\
&L_{ai}=(-1)^{p(a)}\cc_{ai} q^{(-1)^{p(i)}  \n_{i,[i+1,a-1]}}
 \qquad \text{for} \quad i+1 \le a \le M+N, \\
&L_{ib}=(q-q^{-1}) \cd_{ib} q^{(-1)^{p(i)}  \n_{i,[b,i-1]}}
 \qquad \text{for} \quad 1 \le b \le i-1, \\
&L_{ab}=(-1)^{(p(a)+p(b))(p(a)+p(i))+p(i)}
 (q-q^{-1}) \cc_{ai}\cd_{ib} 
  q^{(-1)^{p(i)}  \n_{i,[b,a-1]}} 
 \nonumber \\
 & \qquad \text{for} \quad 1 \le b<a \le i-1
\quad \text{or} \quad i+1 \le b<a \le M+N, \\[8pt]
& \Lo_{\alpha \beta}=0 \quad \text{for} 
\quad \alpha > \beta \quad 
\text{or} \quad 1 \le \alpha \le \beta \le i-1 \quad 
\text{or} \quad i+1 \le \alpha \le \beta \le M+N, \\
& \Lo_{ii}= q^{(-1)^{p(i)} \n_{i,\overline{I}}}, \\
&\Lo_{ai}=(-1)^{p(a)}\cc_{ai} q^{(-1)^{p(i)} 
  (\n_{i,[1,a-1]}+\n_{i,[i+1,M+N]})}
 \qquad \text{for} \quad 1 \le a \le i-1, \\
&\Lo_{ib}=(q-q^{-1}) \cd_{ib} 
  q^{(-1)^{p(i)} ( \n_{i,[1,i-1]} + \n_{i,[b,M+N]}) }
 \qquad \text{for} \quad i+1 \le b \le M+N, \\
& \Lo_{ab}=(-1)^{(p(a)+p(b))(p(a)+p(i))+p(i)}
 (q-q^{-1}) \cc_{ai}\cd_{ib} 
  q^{(-1)^{p(i)}  (\n_{i,[1,a-1]}+\n_{i,[b,M+N]})} 
 \nonumber \\
 & 
\qquad \text{for} \quad 1 \le a <i <  b  \le M+N. 
\end{align} 

{\em (iii) The case}  
$I=\{1,2,\dots, M+N\} \setminus \{ a \}$, 
$\overline{I}=\{ a \}$:  
\begin{align}
& L_{\alpha \beta}=0 \quad \text{for} 
\quad \alpha < \beta,  \\
& L_{aa}= q^{(-1)^{p(a)} \n_{I,a}}, 
\\
&L_{ii}=q^{-(-1)^{p(i)}  \n_{i,a}} 
\qquad \text{for} \quad i \in I, \\
&L_{ia}=(-1)^{p(a)}(q-q^{-1})
 \cd_{ia} q^{(-1)^{p(a)}  (\n_{[1,a-1],a} + \n_{[i+1,M+N],a})}
 \quad \text{for} \quad a+1 \le i \le M+N, \\
&L_{aj}=q^{-(-1)^{p(a)}} \cc_{aj} 
  q^{(-1)^{p(a)} ( \n_{[1,j],a}+\n_{[a+1,M+N],a} )}
 \qquad \text{for} \quad 1 \le j \le a-1, \\
&L_{ij}=(-1)^{(p(i)+p(j))p(a)+p(i)p(j)+1}
 (q-q^{-1}) \cd_{ia}\cc_{aj} 
  q^{-(-1)^{p(a)}  \n_{[j+1,i],a}} 
 \nonumber \\
 & \qquad \text{for} \quad 1 \le j<i \le a-1
\quad \text{or} \quad a+1 \le j<i \le M+N, \\
&L_{ij}=(-1)^{(p(i)+p(j))p(a)+p(i)p(j)}
 q^{-(-1)^{p(a)}}(q-q^{-1}) \cd_{ia}\cc_{aj} 
  q^{(-1)^{p(a)} ( \n_{[1,j],a}+ \n_{[i+1,M+N],a} )} 
 \nonumber \\
 & \qquad \text{for} 
\quad 1 \le j <a <i \le M+N, 
\\[8pt]
& \Lo_{\alpha \beta}=0 \qquad \text{for} 
\quad \alpha > \beta \quad 
\text{or} \quad \alpha=\beta=a, \\
& \Lo_{ii}= q^{(-1)^{p(i)} \n_{i,a}} \qquad \text{for} \quad i \in I, \\
&\Lo_{ia}=(-1)^{p(a)}(q-q^{-1})\cd_{ia} q^{(-1)^{p(a)} 
  \n_{[i+1,a-1],a}}
 \qquad \text{for} \quad 1 \le i \le a-1, \\
&\Lo_{aj}=q^{-(-1)^{p(a)}} \cc_{aj} 
  q^{(-1)^{p(a)} \n_{[a+1,j],a} }
 \qquad \text{for} \quad a+1 \le j \le M+N, \\
& \Lo_{ij}=(-1)^{(p(i)+p(j))p(a)+p(i)p(j)+1}
 (q-q^{-1}) \cd_{ia} \cc_{aj} 
  q^{-(-1)^{p(a)}  (\n_{[1,i],a}+\n_{[j+1,M+N],a})} 
 \nonumber \\
 & 
\qquad \text{for} \quad 1 \le i <a <  j  \le M+N, 
\\ 
& \Lo_{ij}=(-1)^{(p(i)+p(j))p(a)+p(i)p(j)}
  q^{-(-1)^{p(a)}}
 (q-q^{-1}) \cd_{ia} \cc_{aj} 
  q^{(-1)^{p(a)}  \n_{[i+1,j],a}} 
 \nonumber \\
 & 
\qquad \text{for} \quad 1 \le i < j  \le a-1 
\quad \text{or} \quad 
 a+1 \le i < j  \le M+N. 
\end{align} 

{\em (iv) The case} 
 $I=\{1,2,\dots, M+N\}$, $\overline{I}=\emptyset$: 
for $i,j \in I$, 
\begin{align}
& L_{ij}=\Lo_{ij}=0 \quad \text{for} \quad i \ne j \quad 
\text{and} \quad 
L_{ii}=\Lo_{ii}=1. 
\label{L-full}
\end{align}
Expressions of $L_{ij}, \Lo_{ij}$ for  the generic set $I$ 
in terms of the oscillator algebras 
for the case $M+N \ge 4$
are involved especially for $|i-j| \ge 2$, and their explicit forms  
are unknown. 

One may also apply the transformations \eqref{autoosc} or \eqref{autoosc2} to these solutions 
to get many parameter solutions.  
The q-oscillator solutions of the graded Yang-Baxter equation are given by substituting the 
above q-oscillator realizations of the L-operators into the map 
\eqref{rll-ev1}. We denote the corresponding solutions as
\begin{align}
\mathbf{L}_{I}(x)=\mathbf{L}-\mathbf{\overline{L}}x^{-1}.  
 \label{solI}
\end{align} 
We remark that the following renormalized L-operators
\begin{align}
{\mathcal L}_{I}(v):=
1 \otimes 
\left( 
\sum_{i \in I}(q-q^{-1})^{-1}E_{ii} +
\sum_{b \in \overline{I}}E_{bb}
\right)
q^{v}\mathbf{L}_{I}(q^{2v}), 
\quad 
v \in \mathbb{C}
\end{align}
reduce to L-operators similar to the ones in \cite{BFLMS10} in the rational limit $q \to 1$.

Now \eqref{solI} defines an evaluation map from the q-superYangian to the contracted algebra.  
Let us calculate the actions of generators on the vacuum defined by 
\begin{align} 
\n_{ia}|0 \rangle=\cc_{ai}|0\rangle=0 \quad \text{for all} \quad i\in I, a \in \overline{I}. 
 \label{vacosc}
\end{align}
They lead 
\begin{align}
\begin{split} 
&L_{ii}(x) |0 \rangle =(1-x^{-1}) |0 \rangle  \quad  
\quad \text{for}  \quad i \in I, \\
&L_{aa}(x) |0 \rangle = |0 \rangle  \quad  
\quad \text{for} \quad  a \in \overline{I}. 
\end{split}
\label{highestweightos}
\end{align}
In particular for $I=\{1,2,\dots,n \} \subset \{1,2,\dots, M+N \}$,  we find 
\begin{align}
&L_{ij}(x) |0 \rangle =0
 \quad 
\text{for} \quad  i>j. \label{hwcon}  
\end{align}
Thus the corresponding representation is a highest weight representation of 
the q-superYangian with the highest weight vector $|0 \rangle $ and 
the highest weight given by \eqref{highestweightos}. 
In addition, 
the ratio of the eigenvalues $\nu_{i}(x)$ of $L_{ii}(x) $ on $|0 \rangle $ is 
$ \nu_{i}(x)/\nu_{i+1}(x)=1-x^{-1}\delta_{n,i}$ for $1 \le i \le M+N-1$. 
This is a kind of Drinfeld rational fraction\footnote{Here the spectral parameter $x$ came from ${\mathcal B}_{-}$. 
To interpret it as the one from ${\mathcal B}_{+}$, we have to 
replace $x$ with $x^{-1}$.}
 introduced in \cite{HJ11}. 
 The finite dimensional representations of the quantum affine algebras are characterized by the Drinfeld polynomials.
 In contrast, q-oscillator representations given as limits of the Kirillov-Reshetikhin modules
\footnote{The q-characters or the T-functions for the Kirillov-Reshetikhin modules 
solve the T-system for $MN=0$ \cite{KNS93}  and for $MN \ne 0$ \cite{T97}.} of the Borel subalgebra of the quantum 
 affine algebras are characterized by the Drinfeld rational fractions. 
 One may regard \eqref{highestweightos}-\eqref{hwcon} as a new definition of 
 this type of representations in the FRT formulation, which seems to be unknown in the literatures. 
For the other sets $I$, the highest weight condition \eqref{hwcon} will have to be changed since they should be 
 interpreted as representations permuted by automorphisms of $U_{q}(\hat{gl}(M|N))$. 
Let us consider a renormalized L-operator 
\begin{align}
\tilde{\mathbf L}(x)={\mathbf L}(xq^{-2m})(1 \otimes q^{-m \sum_{i \in I} E_{ii}}) 
 \label{renoL}
\end{align}
 for the 
q-superYangian shifted by the automorphisms \eqref{shiftconst} and 
\eqref{shiftL}. 
The latter corresponds to 
\begin{align}
c_{i}=-m \quad  \text{for} \quad  i\in I, 
\qquad  
c_{i}=0 \quad  \text{for} \quad  i\in \overline{I} 
 \label{shiftpara}
\end{align}
 in \eqref{shiftgl}-\eqref{shift-pre-gl}. 
For an evaluation representation based on the map \eqref{rll-ev1} and 
 the highest weight representation of $U_{q}(gl(M|N))$ with the highest weight 
\begin{align}
\lambda_{i} =(-1)^{p(i)} m  \quad \text{for} \quad   i \in I,  \quad 
\text{and} \quad  \lambda_{a} =0  \quad \text{for} \quad  a \in \overline{I} 
\label{highestweight}
\end{align}
(cf.\ \eqref{weightl20}), 
the eigenvalues of the diagonal part of $\tilde{\mathbf L}(x)$ on the highest weight vector 
coincides with the ones in \eqref{highestweightos} in the limit
\footnote{The opposite limit $m \to -\infty $ for $|q|<1 $ (or $m \to \infty $ for $|q|>1 $) [without 
the shift of the spectral parameter in \eqref{renoL}]  will 
 effectively interchange the role of $I$ and $\overline{I}$.} 
$m \to \infty $ for $|q|<1 $ (or $m \to -\infty $ for $|q|>1 $). 
\subsection{Toward contraction of $U_{q}(\hat{gl}(M|N))$}
It will be natural to consider an affine analogue $U_{q}(\hat{gl}(M|N;I))$ 
(or $\tilde{U}_{q}(\hat{gl}(M|N;I))$)  
of $U_{q}(gl(M|N;I))$ (or $\tilde{U}_{q}(gl(M|N;I))$). 
We will discuss how they will look like. 

The evaluation map \eqref{eva} has another presentation of the form:
\begin{align}
\begin{split}
& e_{0} \mapsto -(-1)^{p(1)} x (q-q^{-1})^{-1} \Lo_{1,M+N} L_{M+N,M+N}^{-1}   , \\
& f_{0} \mapsto x^{-1} (q-q^{-1})^{-1} L_{M+N,M+N} L_{M+N,1} , \\
& h_{0} \mapsto \frac{ \log (L_{M+N,M+N} L_{1,1}^{-1}) }{\log q},
\\
& e_{i} \mapsto (-1)^{p(i+1)} (q-q^{-1})^{-1} L_{i+1,i} L_{ii}^{-1}, \\
& f_{i} \mapsto - (q-q^{-1})^{-1} L_{ii} \Lo_{i,i+1} , \\
& h_{i} \mapsto \frac{ \log (L_{ii} L_{i+1,i+1}^{-1}) }{\log q} 
\qquad \text{for} \quad 1 \le i \le M+N-1.
\end{split}
 \label{evslgl}
\end{align}
In addition, the map \eqref{eva-gl} becomes: 
\begin{align}
 k_{i} \mapsto (-1)^{p(i)} \frac{ \log L_{ii}  }{\log q}  
\qquad \text{for} \quad 1 \le i \le M+N. 
\label{evglgl}
\end{align}
We also define
\begin{align}
 \overline{k}_{i} \mapsto (-1)^{p(i)} \frac{ \log \Lo_{ii}  }{\log q}  
\qquad \text{for} \quad 1 \le i \le M+N. 
\label{evglgl2}
\end{align}
Due to the relation \eqref{rll1}, \eqref{evglgl} and \eqref{evglgl2} are consistent with \eqref{kbar}.
 Let us substitute $L_{ij}$ given by \eqref{dia1}-\eqref{L-full} (for a fixed $I$) into the right hand side of 
\eqref{evslgl}-\eqref{evglgl}. 
This gives an evaluation map from ${\mathcal B}_{+}$ or ${\mathcal B}_{-}$ 
 to the q-oscillator superalgebra. We denote this map as $\rho_{I}(x)$. 
 Similar maps from (restricted to) ${\mathcal B}_{+}$  to  the q-oscillator (super)algebra were 
 considered  for $U_{q}(\hat{sl}(2))$ \cite{BLZ97},  $U_{q}(\hat{sl}(3))$ \cite{BHK02}, 
$U_{q}(\hat{sl}(M))$ \cite{Kojima08} and  $U_{q}(\hat{sl}(2|1))$ \cite{BT08}. 
Here we used $L_{ii}^{-1} $ in \eqref{evslgl} instead of $\overline{L}_{ii} $ since $L_{ii}^{-1} $ (for $\in \overline{I} $) 
do not coincide with $\overline{L}_{ii} $ 
for the contracted algebras $U_{q}(gl(M|N;I))$.  
We remark that $\rho_{I}(x)$ is not an evaluation map from 
$U_{q}(\hat{gl}(M|N))$ to the q-oscillator superalgebra but 
rather should be interpreted as a map from a certain contracted algebra
 $U_{q}(\hat{gl}(M|N;I))$ on $U_{q}(\hat{gl}(M|N))$. 
We do not have a rigorous definition of $U_{q}(\hat{gl}(M|N; I))$  in full generality. 
Here we mention relations for $U_{q}(\hat{gl}(M|N; I))$, which we observe through examples. 

First, we find that the following contracted commutation 
relations hold true under the map. 
\begin{align}
[e_{i},f_{j}]= 
\begin{cases}
\delta_{ij} \frac{q^{h_{i}} -q^{-h_{i}} }{q-q^{-1}}  & \text{for} \quad i ,i+1 \in I, 
\\ 
\frac{ \delta_{ij} q^{h_{i}}}{q-q^{-1}}  & \text{for} \quad i \in \overline{I}, 
\quad i+1 \in I, 
 \\
 \frac{-\delta_{ij}  q^{-h_{i}}}{q-q^{-1}}  & \text{for} \quad i \in I, \quad 
i+1 \in \overline{I}, 
\\
 0 & \text{for} \quad i, i+1 \in \overline{I},
\end{cases}
\label{ef-cont}
\end{align} 
where $i,j$ should be interpreted under $ \mod M+N$. 
The other commutation relations  hold true basically in the same way as the ones in section 2. 
However, some of the relations become trivial ($0=0$) when the generator $f_{i}$ vanishes
\footnote{These $f_{i}$ are not original generators of $U_{q}(\hat{gl}(M|N))$ but the limit of renormalized generators of it 
(see Appendix B). 
Original generators $f_{i}$ of $U_{q}(\hat{gl}(M|N))$ can diverge. Then these q-oscillator representations of 
$\mathcal{B}_{+}$ can not be straightforwardly extended to the ones for the whole algebra $U_{q}(\hat{gl}(M|N))$. 
We can still extend them for the contracted algebra $U_{q}(\hat{gl}(M|N;I))$ instead.}:
\begin{align}
f_{i}=0  \qquad \text{for} \quad i,i+1 \in \overline{I}. 
\label{f-zero}
\end{align} 
To be precise, we observed the following non-trivial Serre-type relations in addition to
\eqref{com-ee} and \eqref{Serre-aff}-\eqref{extS3}.
\\
The case $M+N \ge 3$: 
\begin{align}
& [e_{i}, e_{i+1}]_{q^{-a_{i,i+1}}}=[f_{i}, f_{i+1}]_{q^{a_{i,i+1}}}=0 \quad \text{for} \quad i,i+2 \in I, \quad  i+1 \in \overline{I},  
\label{Serre-cont1}
\\
& [e_{i}, e_{i+1}]_{q^{a_{i,i+1}}}=[f_{i}, f_{i+1}]_{q^{-a_{i,i+1}}}=0 \quad \text{for} \quad i,i+2 \in \overline{I}, \quad  i+1 \in I,
\label{Serre-cont2}
\end{align}
where $a_{ij}$ is the Cartan matrix \eqref{Cartan-mat} and the indices  should be interpreted
 under $\mod M+N$. 
\\
The case $(M,N)=(2,0)$ or $(0,2)$: 
\begin{align}
& [e_{0}, [e_{0}, e_{1}]_{q^{a_{01} } } ]=
[e_{1}, [e_{1}, e_{0}]_{q^{-a_{10} } } ]=
[f_{0}, [f_{0}, f_{1}]_{q^{-a_{01} } } ]=
[f_{1}, [f_{1}, f_{0}]_{q^{a_{10} } } ]=0 \nonumber \\
& \qquad \text{for} \quad 1 \in I , \quad  2 \in \overline{I} , 
\label{Serre-cont31}
\\
& [e_{0}, [e_{0}, e_{1}]_{q^{-a_{01} } } ]=
[e_{1}, [e_{1}, e_{0}]_{q^{a_{10} } } ]=
[f_{0}, [f_{0}, f_{1}]_{q^{a_{01} } } ]=
[f_{1}, [f_{1}, f_{0}]_{q^{-a_{10} } } ]=0 \nonumber \\
& \qquad \text{for} \quad 1 \in \overline{I} , \quad  2 \in I. 
\label{Serre-cont32}
\end{align}
The case
\footnote{
For $1 \in \overline{I},  \ 2,3 \in I$ case, 
\eqref{Serre-cont7}  follows 
from \eqref{Serre-cont1};
for $1 \in I,  \ 2,3 \in \overline{I}$ case, 
\eqref{Serre-cont6} follows 
from \eqref{Serre-cont2};
for $2 \in \overline{I},  \ 1,3 \in I$ case, 
\eqref{Serre-cont5}  follows 
from \eqref{Serre-cont1};
for $2 \in I,  \ 1,3 \in \overline{I}$ case, 
\eqref{Serre-cont4}  follows 
from \eqref{Serre-cont2};
for $3 \in \overline{I},  \ 1,2 \in I$ case, 
\eqref{Serre-cont4} (resp.\ \eqref{Serre-cont6}) follows 
from \eqref{Serre-cont1} and  $(e_{2})^2=0$ (resp.\ $(e_{0})^2=0$);
for $3 \in I,  \ 1,2 \in \overline{I}$ case, 
\eqref{Serre-cont5} (resp.\ \eqref{Serre-cont7}) follows 
from \eqref{Serre-cont2} and  $(e_{2})^2=0$ (resp.\ $(e_{0})^2=0$). 
A similar remark can be applied for $(M,N)=(1,2)$ case as well. 
} 
$ (M,N)=(2,1)$: 
\begin{align}
& [e_{2},\,[e_{0},\,[e_{2},\,e_{1}]_{q }]] =[f_{2},\,[f_{0},\,[f_{2},\,f_{1}]_{q^{-1} }]] =0 
\quad \text{for} \quad 1 \in \overline{I},  \quad  2,3 \in I ,
\label{Serre-cont4}
\\
& [e_{2},\,[e_{0},\,[e_{2},\,e_{1}]_{q^{-1} }]] =0 
\quad \text{for} \quad 1 \in I,  \quad  2,3 \in \overline{I} ,
\label{Serre-cont5}
\\
& [e_{0},\,[e_{2},\,[e_{0},\,e_{1}]_{q^{-1}}]]=[f_{0},\,[f_{2},\,[f_{0},\,f_{1}]_{q}]]=0
 \quad \text{for} \quad  2 \in \overline{I}, \quad  1,3 \in I, 
 \label{Serre-cont6}
\\
& [e_{0},\,[e_{2},\,[e_{0},\,e_{1}]_{q}]]=0
 \quad \text{for} \quad  2 \in  I, \quad  1,3 \in \overline{I}. 
\label{Serre-cont7}
\end{align}
The case $ (M,N)=(1,2)$: 
\begin{align}
& [e_{1},\,[e_{0},\,[e_{1},\,e_{2}]_{q }]] =[f_{1},\,[f_{0},\,[f_{1},\,f_{2}]_{q^{-1} }]] =0 
\quad \text{for} \quad 3 \in \overline{I},  \quad  1,2 \in I ,
\label{Serre-cont8}
\\
& [e_{1},\,[e_{0},\,[e_{1},\,e_{2}]_{q^{-1} }]] =0 
\quad \text{for} \quad 3 \in I,  \quad  1,2 \in \overline{I} ,
\label{Serre-cont9}
\\
& [e_{0},\,[e_{1},\,[e_{0},\,e_{2}]_{q^{-1}}]]=[f_{0},\,[f_{1},\,[f_{0},\,f_{2}]_{q}]]=0
 \quad \text{for} \quad  2 \in \overline{I}, \quad  1,3 \in I, 
 \label{Serre-cont10}
\\
& [e_{0},\,[e_{1},\,[e_{0},\,e_{2}]_{q}]]=0
 \quad \text{for} \quad  2 \in  I, \quad  1,3 \in \overline{I}. 
\label{Serre-cont11}
\end{align}
The first equation in \eqref{Serre-cont1} (or \eqref{Serre-cont2}) for $i=0$ and $(M,N)=(3,0)$ case corresponds to 
 the second equation
\footnote{We did not consider the first equation in eq.\  (4.45) in \cite{BHK02}. 
It looks like a statement that the relation is a center  rather than 
it is a Serre-type relation.} in eq.\  (4.45) in \cite{BHK02} (see also \cite{Ridout:2011wx}). 
Some of the Serre-type relations in section 2 automatically hold true under these relations. 
For example, we find the following  relations
\footnote{
If we can relax the conditions on the indices for \eqref{Serre-cont12}-\eqref{Serre-cont13} 
(in particular, if we can drop the condition $i+3 \in I$ in \eqref{Serre-cont12}, and 
 the condition $i+3 \in \overline{I}$ in \eqref{Serre-cont13}), then these become independent of the relations 
 \eqref{Serre-cont1} and  \eqref{Serre-cont2}. 
}: 
\begin{align}
 & [[e_{i}, e_{i+1}]_{q^{-a_{i,i+1} } } , e_{i+2}]_{q^{-a_{i+1,i+2}} } =[[f_{i}, f_{i+1}]_{q^{a_{i,i+1} } } , f_{i+2}]_{q^{a_{i+1,i+2}} }
 =0 \nonumber 
\\
& \qquad \qquad \text{for} \quad  i,i+1,i+3 \in I,  \quad   i+2 \in \overline{I}, \quad  M+N \ge 4 , 
\label{Serre-cont12} 
 \\
&  [[e_{i}, e_{i+1}]_{q^{a_{i,i+1} } } , e_{i+2}]_{q^{a_{i+1,i+2}} } 
 =0  \nonumber 
\\
& \qquad \qquad \text{for} \quad  i,i+1,i+3 \in \overline{I},  \quad  i+2 \in I, \quad M+N \ge 4 .
\label{Serre-cont13} 
\end{align}
\eqref{Serre-cont12} (resp.\ \eqref{Serre-cont13} ) follow from 
\eqref{com-ee}  and \eqref{Serre-cont1} (resp.\ \eqref{Serre-cont2}). 
Then, \eqref{extraS} follow from these if $i=M-1$ or $M+N-1$.  
Note that these relations \eqref{Serre-cont1}-\eqref{Serre-cont7} are not symmetric  
under $q \leftrightarrow q^{-1}$, although the original Serre-type relations in section 2 are symmetric 
 under this. 

Our L-operators \eqref{solI} satisfy the defining relations of the universal R-matrix. 
In particular, the following relations are valid 
\begin{multline}
 \left( 1 \otimes \pi(y)( k_{i} ) + \rho_{I}(x)(k_{i}) \otimes 1 \right) \Lf_{I}(y/x) 
= \\
=
\Lf_{I}(y/x)  \left( 
 \rho_{I}(x)( k_{i} ) \otimes 1  + 1 \otimes \pi(y)(k_{i} ) 
\right) ,
\label{intert1}
\end{multline}
\begin{multline}
 \left( 1 \otimes \pi(y)( e_{i} ) + \rho_{I}(x)(e_{i}) \otimes \pi(y) (q^{-h_{i}}) \right) \Lf_{I}(y/x) 
= \\
=
\Lf_{I}(y/x)  \left( 
 \rho_{I}(x)( e_{i} ) \otimes 1  + \rho_{I}(x) (q^{-h_{i}}) \otimes \pi(y)(e_{i} ) 
\right) ,
\label{intert2}
\end{multline}
where $ 0 \le i \le M+N-1$ ($k_{0}=k_{M+N}$). This is because 
 our L-operators are image of the universal R-matrix (up to an overall factor $N_{I}(x,y) $): 
$\Lf_{I}(y/x)= N_{I}(x,y) ( \rho_{I}(x) \otimes \pi (y) ) (\tilde{\mathcal R}) $ 
(see also discussions on the universal R-matrix in \cite{Ridout:2011wx}). 
Note that the relation for $f_{i}$, namely 
\begin{multline}
 \left( \rho_{I}(x) (q^{h_{i}})  \otimes \pi(y)( f_{i} )  \theta(i+1 \in I)
+ \rho_{I}(x)(f_{i}) \otimes 1 \right) \Lf_{I}(y/x) 
= \\
=
\Lf_{I}(y/x)  \left( 
 \rho_{I}(x)( f_{i} ) \otimes \pi(y)(q^{h_{i}} )  +
 1 \otimes \pi(y)(f_{i} ) \theta(i \in I)
\right) 
\label{intert3}
\end{multline}
has the standard form
 only for the case $ i ,i+1 \in I$ ($0 \equiv M+N $) since we are considering a contracted algebra $U_{q}(\hat{gl}(M|N;I))$. 
 In particular, this can be $0=0$ for  $ i ,i+1 \notin I$ case. 

We have observed the relations \eqref{ef-cont}-\eqref{intert3} under the map $\rho_{I}(x)$. 
To be precise, 
\eqref{ef-cont} follows from the maps \eqref{evslgl}-\eqref{evglgl2} for \eqref{efk} with the 
contraction \eqref{red2}. 
\eqref{f-zero} follows from the map \eqref{evslgl} with the contraction \eqref{red30}-\eqref{red3}. 
\eqref{Serre-cont31}-\eqref{Serre-cont32} follow from the map \eqref{evslgl} with 
the contraction \eqref{red2}. 
Thus, the map is an algebra homomorphism. 
However, \eqref{Serre-cont1}-\eqref{Serre-cont2} and 
\eqref{Serre-cont4}-\eqref{Serre-cont11} seem to be true only under the map $\rho_{I}(x)$, 
and thus can be representation theoretical relations rather than 
algebraic relations.

Now we want to consider these 
 from an opposite direction. 
 Namely, we may interpret some of the relations 
 \eqref{ef-cont}-\eqref{Serre-cont13}, 
\eqref{com-ee}-\eqref{extS3},  
and \eqref{hatglmm} as the defining relations of the contracted algebras 
 $\tilde{U}_{q}(\hat{gl}(M|N;I))$ and 
$U_{q}(\hat{gl}(M|N;I))$. There is a certain arbitrariness on which relations 
should be included in the defining relations. Apparently, \eqref{ef-cont} 
(resp.\ \eqref{f-zero}) is a consequence of an affine analogue 
of the contraction \eqref{red2} 
 (resp.\ subsidiary contraction \eqref{red3}). 
 Then we propose to include 
 \eqref{ef-cont},   
\eqref{com-ee}-\eqref{extS3} 
and \eqref{hatglmm} 
in the defining relations of $\tilde{U}_{q}(\hat{gl}(M|N;I))$; 
and 
\eqref{ef-cont}-\eqref{f-zero}, 
 \eqref{Serre-cont31}-\eqref{Serre-cont32},  
\eqref{com-ee}-\eqref{Serre-aff}, \eqref{extraS}-\eqref{extS3} 
and \eqref{hatglmm} 
in the defining relations of $U_{q}(\hat{gl}(M|N;I))$. 
We expect these fix the whole contracted algebras for the case 
$|\overline{I}|=1$. However, we may have to add more generators and 
relations
\footnote{
This can be guessed from an example on the 
finite algebra $U_{q}(gl(M|N;I))$. 
For  $U_{q}(gl(M|N))$, the generator $\Lo_{ij}$ ($ |i-j| \ge 2$) can be fixed by 
the relation \eqref{relationglmn7} and the generators 
$ \Lo_{k,k+1}, \Lo_{k+1,k+1}$ ($i \le k \le j-1$), which are directly related to 
the Chevalley type generators. 
However this is not always the case for $U_{q}(gl(M|N;I))$ since  
the relation \eqref{relationglmn7} can be trivial (from $\Lo_{j-1,j}=0$, $\Lo_{j,j}=0$) while 
$\Lo_{ij}$ is not for $i \in I$, $j-1,j \in \overline{I} $.    
Then the Chevalley type generators may not be enough to fix the whole contracted algebra 
(explicit relations among  $\Lo_{ij},L_{ij}$ ($ |i-j| \ge 2$) may be necessary).}
 for the case $|\overline{I}| \ge 2 $. 
The restriction of the generators to $\{ e_{i},f_{i} \}_{i=1}^{M+N-1}$ and $\{ k_{i} \}_{i=1}^{M+N}$
 gives relations of $U_{q}(gl(M|N;I))$. 
Then we can consider evaluation representations of $U_{q}(\hat{gl}(M|N;I))$ based on the representations of $U_{q}(gl(M|N;I))$. 
The co-product
\footnote{This `co-product' is different from the usual one in that 
 ${\mathcal  A}$ and ${\mathcal  B}$ for  
$\Delta: {\mathcal  A} \mapsto  {\mathcal  A} \otimes {\mathcal  B} $ 
are different algebras.}
 $\Delta: U_{q}(\hat{gl}(M|N;I)) \mapsto U_{q}(\hat{gl}(M|N;I)) \otimes U_{q}(\hat{gl}(M|N))$ 
for $e_{i}$ and $k_{i}$ is the same as the one in section 2, while the one for $f_{i}$  is (as observed from \eqref{intert3})
 contracted as
\begin{align}
\Delta(f_{i})& = f_{i}  \otimes q^{h_{i}} +  \theta(i \in I) (1 \otimes f_{i}) , \\
\Delta^{\prime}(f_{i}) & = \theta(i+1 \in I) (q^{h_{i}}  \otimes  f_{i} )  
+ f_{i} \otimes 1.
\end{align}
This may be rewritten as 
\begin{align}
\Delta(f_{i}) = f_{i}  \otimes q^{(-1)^{p(i)}k_{i} +(-1)^{p(i+1)}\overline{k}_{i+1} } + 
q^{(-1)^{p(i)}k_{i} +(-1)^{p(i)}\overline{k}_{i} } \otimes f_{i} 
\end{align}
since\footnote{Here \eqref{kbar} is not always true since the generators are renormalized.}
\begin{align}
q^{(-1)^{p(i)}\overline{k}_{i}} =
\begin{cases}
 \theta(i \in I) q^{-(-1)^{p(i)}k_{i}} & \text{for} \quad  U_{q}(\hat{gl}(M|N;I)) \\
 q^{-(-1)^{p(i)}k_{i}} & \text{for} \quad  U_{q}(\hat{gl}(M|N)).
\end{cases}
\end{align}
The co-product $\Delta(\overline{k}_{i})= \overline{k}_{i} \otimes 1 +  1 \otimes \overline{k}_{i} $ is well defined only for $i \in I $ since
 $ \overline{k}_{i} \in U_{q}(\hat{gl}(M|N;I))  $  diverges for $i \in \overline{I}$. However 
 $\Delta(q^{\overline{k}_{i}})= q^{\overline{k}_{i} }\otimes q^{ \overline{k}_{i}} $ is still well defined even for $i \in \overline{I}$ (it just becomes $0$).  

 We may be able to define contracted universal R-matrices in $U_{q}(\hat{gl}(M|N;I)) \otimes {\mathcal B}_{-}$ 
by the contracted co-products for the contracted algebras and
 \eqref{R-def}. 
They will be the universal R-matrices for the Q-operators. 
Of course, 
the existence of such an object is not a trivial issue. 
The universal R-matrix for $U_{q}(\hat{gl}(M|N))$ is a sort of 
a power series of the generators of $U_{q}(\hat{gl}(M|N))$. 
The generators of $U_{q}(\hat{gl}(M|N;I))$ are considered to be reductions of 
the generators of $U_{q}(\hat{gl}(M|N))$. Thus, the universal R-matrix for  
$U_{q}(\hat{gl}(M|N; I))$ may be a reduction of the universal R-matrix 
 for $U_{q}(\hat{gl}(M|N))$ as a power series on the generators  (up to the normalization). 
More formally, this may be shown by realizing 
$U_{q}(\hat{gl}(M|N;I))$ as a kind of Drinfeld double
\footnote{We thank a referee for this comment.}.
 Furthermore, it will be important to construct and evaluate a contracted universal R-matrix in $U_{q}(\hat{gl}(M|N;I)) \otimes U_{q}(\hat{gl}(M|N; J))$. 
For this, we may have to repeat similar calculations discussed in the appendix B for $ {\mathcal B}_{+}$ as well as $ {\mathcal B}_{-}$. 
The original universal R-matrix (under a certain condition) may be factorized with respect to contracted universal R-matrices. 
This could be a step toward the construction of the Q-operators for the 
generic representations on the quantum space. 

We may also interpret $U_{q}(\hat{gl}(M|N;I)) $ as a subalgebra of an asymptotic algebra (cf.\ \cite{HJ11}) associated with $U_{q}(\hat{gl}(M|N)) $. 
In terms of the asymptotic algebra, the vanishing of the action of the Cartan generator $ q^{ \overline{k}_{i}} $ for $i \in \overline{I}$ in 
\eqref{efk} 
occurs on the 
level of the representation. Here we regarded this as a phenomenon on the level of the algebra and discussed  the contracted algebra $U_{q}(\hat{gl}(M|N;I)) $. 

As for the FRT formulation of $U_{q}(\hat{gl}(M|N;I)) $, we will have to replace the condition 
\eqref{rlla2} with 
\begin{align}
& L^{(0)}_{ii} \Lo^{(0)}_{ii}=\Lo^{(0)}_{ii}L^{(0)}_{ii}=1
\quad \text{for} \quad i \in I, 
\\
& \Lo^{(0)}_{ii}=0 \qquad \text{for} \quad i \in \overline{I} . 
\label{vanish-L0}
\end{align}
On the other hand, in the context of the asymptotic algebra, we just forget about \eqref{rlla2} and interpret that \eqref{vanish-L0} 
occurs on the level of the representation. 

In this paper, we consider contractions defined by \eqref{red1}-\eqref{red2}. 
Instead of \eqref{red2}, one can consider the following:
\begin{align}
L_{ii}&=0
 \quad \text{for} \quad i \in \overline{I}. 
\label{red4}
\end{align}
The L-operators based on this contraction have one to one correspondence to the ones 
proposed in this paper. They seem to be the image of the Cartan anti-involution for our L-operators. 
One may also consider more general contractions than \eqref{red2} and \eqref{red4}:
\begin{align}
L_{ii}&=0 \quad \text{for} \quad i \in \overline{I}_{1},
\qquad 
\overline{L}_{ii}=0 \quad \text{for} \quad i \in \overline{I}_{2}, 
\qquad 
 \overline{I}_{1},  \overline{I}_{2} \subset  \overline{I}.
\label{red5}
\end{align}
This defines more degenerated algebras and gives degenerated 
solutions of the graded Yang-Baxter equation. 
\section{T- and Q-operators}
In this section, we define Q-operators based on the q-oscillator representations introduced in the previous section and 
 sketch an idea how to write the T-operators in terms of them. 
 This gives a cue for operator realization of the formulas in our previous papers \cite{T09,Tsuboi:2011iz}. 
 
We introduce the universal boundary operator
\begin{align} 
{\mathcal D}=q^{\sum_{i=1}^{M+N} \varphi_{i} k_{i} },  
 \label{boundary}
\end{align}
where $\varphi_{i} \in {\mathbb C}$. 
This boundary operator is a Cartan element of
 $U_{q}(\hat{gl}(M|N))$. Due to the first relation in \eqref{R-def}, 
its co-product commutates with the universal R-matrix 
\begin{align}
 \tilde{\mathcal R} ( {\mathcal D} \otimes {\mathcal D})=
 ({\mathcal D} \otimes {\mathcal D}) \tilde{\mathcal R}. 
 \label{RD}
\end{align}
The images of the evaluation map 
\eqref{eva} and $\rho_{I}(x)$
are given as
\begin{align}
&{\mathbf D}:=\mathsf{ev}_{x}( {\mathcal D}) 
= q^{\sum_{i=1}^{M+N} \varphi_{i} e_{ii} }, 
\\
& {\mathbf D}_{I}:=\mathsf{\rho}_{I}(x)( {\mathcal D}) 
= q^{\sum_{i \in I,a\in \overline{I}} 
(\varphi_{i}- \varphi_{a})\n_{ia} }.
\end{align}
We define the universal T-operator by 
\begin{align}
\Tb_{\lambda}(x)=(\mathrm{Str}_{\pi_{\lambda}(x)} \otimes 1)  
\left[ \tilde{\mathcal R} ( {\mathcal D} \otimes 1) \right]. 
 \label{Top}
\end{align} 
Note that $\Tb_{\lambda}(x)$ is an element of ${\mathcal B}_{-}$ and this definition of the T-operator does not depend on the particular representation of the 
quantum space. 
It is convenient to introduce operators
\begin{align}
z_{i}=q^{(-1)^{p(i)} k_{i} + \varphi_{i} },  
 \label{twistz}
\end{align}
where $1 \le k \le M+N$. 
Then the T-operator \eqref{Top} can be rewritten as
\begin{align}
\Tb_{\lambda}(x)=(\mathrm{Str}_{\pi_{\lambda}(x)} \otimes 1)  
\left[  \overline{ \mathcal R} 
\, 
\overline{\mathcal D}
\right], 
 \label{uni-T}
\end{align}
where 
\begin{align}
\overline{\mathcal D} := q^{\tilde{\mathcal{K}}}( {\mathcal D} \otimes 1)
=
\prod_{j=1}^{M+N} 
\left(
 1 \otimes z_{j} 
\right)
^{k_{j} \otimes 1},
\label{boundary-ren}
\end{align}
where $\tilde{\mathcal{K}}$ is introduced in \eqref{pre-R-re}. 
Here we have renormalized the boundary operator \eqref{boundary}
by the prefactor of the universal R-matrix \eqref{R-red} 
as in \cite{BT08}. 

In the $U_{q}(\hat{sl}(M|N))$-picture, we may define \eqref{boundary}, \eqref{twistz} and 
\eqref{boundary-ren} respectively as
\begin{align}
 {\mathcal D}&=q^{\sum_{k=1}^{M+N-1}\sum_{i=1}^{k}(-1)^{p(i)}\varphi_{i} h_{k} }, 
\qquad 
z_{k}=q^{\varphi_{k}+(-1)^{p(k)} \sum_{j=1}^{M+N-1}(d_{kj}-d_{k-1,j}) h_{j} } ,
\\
\overline{\mathcal D} &:= q^{\mathcal{K}}( {\mathcal D} \otimes 1)
=
\prod_{k=1}^{M+N-1} 
\left(
\prod_{i=1}^{k} (1 \otimes z_{i}^{(-1)^{p(i)}})
\right)
^{h_{k} \otimes 1} , 
\end{align}
where $d_{k0}=d_{M+N,j}=0$, and the parameter $\varphi_{M+N}$ is defined by the relation 
  $\sum_{i=1}^{M+N} (-1)^{p(i)}\varphi_{i} =0$.
In this case, the following relation holds: $\prod_{k=1}^{M+N}z_{k}^{(-1)^{p(k)}}=1$. 

If there is no reduced universal R-matrix in \eqref{uni-T}, 
the following quantity 
\begin{align}
{\mathbb Z}(\lambda)=(\mathrm{Str}_{\pi_{\lambda }(x)} \otimes 1)  
\left[  \overline{\mathcal D} \right], 
 \label{superch}
\end{align} 
gives the supercharacter. For finite dimensional modules, it is a supersymmetric Schur function on the variables \eqref{twistz}. 
In particular for the Verma module, it leads
\begin{align}
& {\mathbb Z}^{+}(\lambda) :=(\mathrm{Str}_{\pi_{\lambda }^{+}(x)} \otimes 1)  
\left[  \overline{\mathcal D} \right] 
=\frac{\prod_{j=1}^{M}  z_{j}^{\lambda_{j} + M-N-j} 
\prod_{k=M+1}^{M+N}  (-z_{k})^{\lambda_{k} + N+M-k}  }{{\mathsf D}} ,
 \label{superchVerma} 
\\
&  \qquad 
{\mathsf D}:= \frac{\prod_{1 \le b < b^{\prime} \le M} (z_{b} - z_{b^{\prime} } ) 
\prod_{M+1 \le f < f^{\prime} \le M+N} (z_{f^{\prime} } -z_{f} )  }
{\prod_{b=1}^{M} \prod_{f=M+1}^{M+N} (z_{b} - z_{f } ) } .
 \label{superVan}
\end{align}
In the above formulas, the reduced universal R-matrix 
plays a role to put the spectral parameter into the supercharacters, 
or to change the  supercharacters to the q-supercharacters. 
This induces sort of shits on the parameters \eqref{twistz} in the supercharacters. 
Let ${\mathcal F}_{I} $ be the Fock space defined by the action of the generators 
$\{ \cc_{ai}, \cd_{ia}, \n_{ia} \}$
 ($i \in I$, $a \in \overline{I}$) of  the q-oscillator superalgebras on the 
vacuum \eqref{vacosc}. 
We define the universal Q-operator by 
\begin{align}
\Qb_{I}(x)=\mathbb{Z}_{I}^{-1}(\mathrm{Str}_{{\mathcal F}_{I}}\otimes 1)  
(\rho_{I}(x) \otimes 1) 
\left[ \overline{\mathcal R} \, \overline{\mathcal D} \right], 
 \label{Qop}
\end{align} 
where the normalization function reads 
\begin{align}
\mathbb{Z}_{I}=(\mathrm{Str}_{{\mathcal F}_{I}} \otimes 1)  
(\rho_{I}(x) \otimes 1) 
\left[  \overline{\mathcal D}  \right].
 \label{norQ}
\end{align} 
Note that these are elements of ${\mathcal B}_{-}$.  
We remark that \eqref{Qop} is 
basically fixed by the map $\rho_{I}(x)$ and the defining relations of the 
q-oscillator superalgebra \eqref{qosc} and does not depend on the definition of the vacuum  (see section 5.2.3 in \cite{BT08} 
for more details). 
Due to the commutation relation \eqref{RD} and \eqref{YBE}, 
the universal T- and Q-operators are commutative
\footnote{To prove the commutativity of the Q-operators algebraically, we need \eqref{YBE} for the contracted universal R-matrix 
 in $U_{q}(gl(M|N; I)) \otimes U_{q}(gl(M|N; J))$, which we do not discuss in this paper. 
 Or, one may prove this on the level of the representation
 (an isomorphism between the tensor product of two auxiliary spaces).}.
\begin{align} 
\begin{split}
 & \Tb_{\lambda}(x)\Tb_{\mu}(y)=\Tb_{\mu}(y) \Tb_{\lambda}(x), 
\qquad 
\Tb_{\lambda}(x) \Qb_{I}(y) =\Qb_{I}(y) \Tb_{\lambda}(x), 
\\
& \Qb_{I}(x) \Qb_{J}(y) = \Qb_{J}(y) \Qb_{I}(x) ,
\end{split}
\end{align}
where $x,y \in {\mathbb C}$, $I,J \subset \{1,2,\dots, M+N \}$ and $\lambda,\mu$ are any highest weights. 

Let us calculate the supertrace \eqref{norQ} over the Fock space ${\mathcal F}_{I}$. 
Explicitly, it leads 
\begin{align}
 \mathbb{Z}_{I}=\prod_{i \in I}\prod_{a \in \overline{I}} 
\left(
1-\frac{z_{a}}{z_{i}}
\right)^{-(-1)^{p(i)+p(a)}} . 
 \label{noros}
\end{align}
As expected, this coincides with a limit of 
a normalized character of 
the Kirillov-Reshetikhin module at least for the case
\footnote{We have also checked that a normalized 
 Sergeev-Pragacz formula 
produces \eqref{noros} in the large Young diagram limit 
under a similar condition 
for the case $MN \ne 0$.} 
$N=0$ (cf. \cite{HJ11}):. 
\begin{multline} 
 \mathbb{Z}_{I}= \lim_{m \to \infty} \frac{\mathrm{S}_{\lambda}(z_{1},z_{2},\dots, z_{M})}{\prod_{k=1}^{M}z_{k}^{\lambda_{k}}} ,
\qquad |z_{i}| > |z_{a}| 
 \quad \text{for all} \quad i\in I, a \in \overline{I}, \\
m:=\lambda_{k} \quad \text{for} \quad k \in I, 
\qquad \lambda_{k}=0 
\quad \text{for} \quad k \in \overline{I}, 
 \label{limit-KR}
\end{multline}
where $\mathrm{S}_{\lambda}(z_{1},z_{2},\dots, z_{M})
=\det_{1 \le i,j \le M}(z_{i}^{M+\lambda_{j}-j})/
\det_{1 \le i,j \le M}(z_{i}^{M-j})$ 
is the Schur function. 
Here we meant the equality in \eqref{limit-KR} by the substitution 
of elements of ${\mathcal B}_{-}$ \eqref{twistz} for the 
complex numbers $\{z_{k}\}$ on the right hand side after the limit. 
The normalization factor in \eqref{limit-KR} came from the shift automorphism 
  \eqref{shiftgl} on $\mathcal{B}_{+}$ for the parameters in \eqref{shiftpara}. 
We expect \cite{T09,Tsuboi:2011iz} that the T-operator is given by 
the Baxterizaiton of the supercharacter
\footnote{
\label{footshift}
The shift of the spectral parameter of the Q-operators in 
\cite{T09,Tsuboi:2011iz} 
can be recovered by putting $q \to q^{-1}$ after the replacement 
$ \Qb_{I }(x) \mapsto \Qb_{I } (xq^{\sum_{k \in I} (-1)^{p(k)}})$.}
\begin{align}
\Tb_{\lambda}(x)=\frac{1}{\Ds} \prod_{k=1}^{M+N} 
\Qb_{\{ k \}} (xq^{-d_{k} -2(N-M+(-1)^{p(k)}) }) \cdot 
\left[\Ds \, {\mathbb Z}(\lambda) \right] 
 \label{baxterize} 
\end{align}
where $d_{k}$ are differential operators which evaluate the degrees of the 
monomials on $\{z_{j}\}$ in the right of the dot $\cdot $. 
They effectively 
 act as $d_{k}=2(-1)^{p(k)} z_{k} \frac{\partial }{\partial z_{k}} $ in $[\cdots ]$.  
 We assume $d_{k}$ act on the functions in the left of the dot $\cdot $ as 
just an identity, although $\{ \Qb_{ \{k\} }\}$ are also functions of $\{z_{k}\}$.
In particular for the Verma module
\footnote{This formula 
 \eqref{VermaT} 
was presented first as a poster at a conference 
`Integrability in Gauge and String Theory 2010', 
Nordita, Sweden, 28 June 2010 - 2 July. 
To fit the formula in \cite{Tsuboi:2011iz}, one has to 
make an overall shift of the spectral parameter $x \to xq^{2(M-N)}$ 
(in the right hand side of \eqref{VermaT}) 
after the manipulation in the footnote \ref{footshift}.}, 
we have \cite{Tsuboi:2011iz}
\begin{align}
\Tb_{\lambda}^{+}(x)=\mathbb{Z}^{+}(\lambda )
 \prod_{j=1}^{M+N} \Qb_{\{ j \}} (xq^{-2\left(
(-1)^{p(j)} \lambda_{j} - \sum_{k=1}^{j-1} (-1)^{p(k)} \right)}) .
\label{VermaT}
\end{align}
We remark that the most of the T-operators can be written as 
summentions of the above formula \eqref{VermaT} based on 
the Bernstein-Gelfand-Gelfand resolution and rewritten as  Wronskian-like determinants 
(see \cite{BLZ97} for $U_{q}(\hat{sl}(2))$, \cite{BHK02} for $U_{q}(\hat{sl}(3))$, 
\cite{Kojima08} for finite dimensional representations of $U_{q}(\hat{sl}(M))$ (see also a Wronskian like determinant in 
\cite{KLWZ97}), 
\cite{BT08} for $U_{q}(\hat{sl}(2|1))$; 
\cite{T09,Tsuboi:2011iz} for the Wronskian-like determinants for any $U_{q}(\hat{gl}(M|N))$). 
We expect our universal Q-operators obey functional relations of the form:
for $p(i)=p(j)$:
\begin{multline}
 (z_{i}-z_{j}) \Qb_{I}(xq^{1-2p(i)}) \Qb_{I \cup \{i,j\}}(xq^{-1+2p(i)})=\\
=
z_{i} \Qb_{I \cup \{i\}}(xq^{-1+2p(i)}) \Qb_{I \cup \{j\}}(xq^{1-2p(i)})-
z_{j} \Qb_{I\cup \{i\}}(xq^{1-2p(i)}) \Qb_{I \cup \{j\}}(xq^{-1+2p(i)}), 
\label{QQ1}
\end{multline}
and for $p(i) \ne p(j)$:
\begin{multline}
 (z_{i}-z_{j}) \Qb_{I\cup \{i\}}(xq^{-1+2p(i)}) \Qb_{I\cup \{j\}}(xq^{1-2p(i)})=
\\
=
z_{i} \Qb_{I}(xq^{1-2p(i)}) \Qb_{I\cup \{i,j\}}(xq^{-1+2p(i)})-
z_{j} \Qb_{I}(xq^{-1+2p(i)}) \Qb_{I\cup \{i,j\}}(xq^{1-2p(i)}) .
\label{QQ2}
\end{multline}
At the moment, 
these functional relations are fully proven 
 for  $U_{q}(\hat{sl}(2))$ \cite{BLZ97},  for $U_{q}(\hat{sl}(3))$ \cite{BHK02} and 
 for $U_{q}(\hat{sl}(2|1))$ \cite{BT08}. Their proof is based on decompositions  
 of q-oscillator representations of ${\mathcal B}_{+}$ and does not rely on the 
 representation of  ${\mathcal B}_{-}$ on the quantum space. 
 See also \cite{Kazakov:2010iu,BFLMS10} for discussions on rational models ($q=1$). 
 On the level of the eigenvalues of Q-operators for rational models, \eqref{QQ2} were discussed in details 
in relation to the B\"{a}cklund transformations \cite{Kazakov:2007fy}. 
Here we used expressions based on the $2^{M+N}$ index sets on the Hasse diagram presented in \cite{T09}. 

Now that we have the universal T-and Q-operators \eqref{Top}, \eqref{Qop}, our next task is to evaluate these for 
particular representations of ${\mathcal B}_{-}$ 
on the quantum space of the model.  
For example, the T-operator for the lattice model 
whose quantum space is the fundamental representation on 
each site is given as
\begin{align}
\Tf_{\lambda} (x)&=
N_{\lambda}^{(L)}(x)
\left(
\pi(\xi_{1}) \otimes \pi(\xi_{2}) \cdots \otimes \pi(\xi_{L}) 
\right)
 \left[ \Delta^{(L-1)} \Tb_{\lambda}(x) \right] \\
&= \mathrm{Str}_{\pi_{\lambda}}
\left[
 \Lf^{0L}(\xi_{L}/x)\cdots 
\Lf^{02}(\xi_{2}/x) \Lf^{01}(\xi_{1}/x) ({\mathbf D} \otimes 1^{\otimes L})
\right], 
 \label{latticeT}
\end{align}
where $L$ is the number of the lattice site;  
the complex parameters $\{\xi_{j}\}_{j=1}^{L}$ are inhomogeneities on the spectral parameter; and 
$N_{\lambda}^{(L)}(x)$ is a function for the normalization.  
In \eqref{latticeT}, the evaluation map \eqref{rll-ev1} is used and 
the supertrace is taken over the auxiliary space denoted as `$0$'.
The Q-operators for the same system are given by 
\begin{align}
\Qf_{I} (x) &=
N_{I}^{(L)}(x)
\left(
\pi(\xi_{1}) \otimes \pi(\xi_{2}) \cdots \otimes \pi(\xi_{L}) 
\right)
 \left[ \Delta^{(L-1)} \Qb_{I}(x) \right] \\
&= 
\mathbf{Z}_{I}^{-1}
\mathrm{Str}_{\mathcal{F}_{I}}
\left[
 \Lf^{0L}_{I}(\xi_{L}/x)\cdots 
\Lf^{02}_{I}(\xi_{2}/x) \Lf^{01}_{I}(\xi_{1}/x)
 ({\mathbf D}_{I} \otimes 1^{\otimes L})
\right], 
\label{latticeQ}
\end{align}
where $\mathbf{Z}_{I} :=  \left( \pi(\xi_{1}) \otimes \pi(\xi_{2}) \cdots \otimes \pi(\xi_{L}) \right)  
\left[ \Delta^{(L-1)} \mathbb{Z}_{I} \right]$ and 
the normalization function is 
$N_{I}^{(L)}(x):=\prod_{k=1}^{L}N_{I }(x,\xi_{k})$. 
It is instructive to calculate the lattice T-operator \eqref{latticeT} for the Verma module
\footnote{We remark that a formula similar to the first equality in 
\eqref{1siteT} (for the characters of 
 finite dimensional representations of $U_{q}(gl(M))$) was previously 
derived by Anton Zabrodin in 2007 based on the trigonometric version of 
 the co-derivative for $L=1$ case.} and 
the lattice Q-operator 
\eqref{latticeQ}
even for one site $L=1$ case.  Let us introduce a notation 
$\mathbf{Z}^{+}(\lambda ):=\pi(\xi_{1}) (\mathbb{Z}^{+}(\lambda ) )$. Then we obtain
\begin{align}
\begin{split}
& \left[ \Tf_{\lambda}^{+} (x)\right]_{ii} 
 =\mathbf{Z}^{+}( \lambda )-\frac{x}{\xi_{1}} q^{-d_{i}}  \cdot
\mathbf{Z}^{+}(\lambda ) = 
 \\
& =
\mathbf{Z}^{+}(\lambda )
 \left(
1-\frac{x 
q^{-2 
 \left( 
  (-1)^{p(i)} \lambda_{i} -\sum_{k=1}^{i-1}(-1)^{p(k)}  
  \right)
  }
}{\xi_{1}} 
\prod_{b=1 \atop b \ne i}^{M+N} 
\left(
 \frac{1-\frac{z_{b}}{z_{i}}}
{1-\frac{z_{b}q^{2(-1)^{p(i)}}}{z_{i}}}
\right)^{(-1)^{p(i)+p(b)}} 
\right)
  \\
 & \hspace{101pt}  \text{for} \quad 1 \le i \le M+N ,
 \\
&
\left[ \Tf^{+}_{\lambda}(x) \right]_{\alpha \beta} =0
 \qquad \text{for} \quad  \alpha \ne \beta,
\end{split}
 \label{1siteT} 
\end{align}
\begin{align}
\begin{split}
\left[ \Qf_{I} (x)\right]_{ii}&=1-\frac{x}{\xi_{1}} \frac{q^{-d_{i}} \cdot 
 \mathbf{Z}_{I} }{\mathbf{Z}_{I} }=
1-\frac{x}{\xi_{1}} \prod_{b \in \overline{I}} 
\left(
 \frac{1-\frac{z_{b}}{z_{i}}}
{1-\frac{z_{b}q^{2(-1)^{p(i)}}}{z_{i}}}
\right)^{(-1)^{p(i)+p(b)}} 
 \, \text{for} \quad i \in I, 
\\
\left[ \Qf_{I} (x)\right]_{aa}&=1 
 \quad \text{for} \quad a \in \overline{I}, 
\\
\left[ \Qf_{I} (x)\right]_{\alpha \beta}&=0
 \quad \text{for} \quad  \alpha \ne \beta,
\end{split}
 \label{1siteQ}
\end{align}
where $\left[ {\mathbf M} \right]_{\alpha \beta}$ denotes 
the $(\alpha, \beta)$ matrix element of a ($(M+N)\times (M+N)$) matrix $ {\mathbf M} $.
 In \eqref{1siteT} and \eqref{1siteQ}, 
 the twist parameters should be interpreted as $(i,i)$-matrix element of them
\footnote{In the $U_{q}(\hat{sl}(M|N)) $ picture, this is 
$z_{k}=[z_{k}]_{ii}=
q^{\varphi_{k} +(-1)^{p(k)+p(i)}
(d_{ki} -d_{k-1,i}-d_{k,i-1}+d_{k-1,i-1} )} . 
$
}:
\begin{align}
z_{k}=[z_{k}]_{ii}=
q^{\varphi_{k} +(-1)^{p(k)}\delta_{ik} } . 
\end{align}
The above example gives a non-trivial support to the 
QQ-relations \eqref{QQ1}-\eqref{QQ2} and 
the factorization formulas \eqref{VermaT} for the Verma module 
as the shape of these equations will be essentially independent of the quantum space
 of the model. 
This also agrees with examples in eqs. (3.38)-(3.43) in \cite{BT08} 
up to a transformation $q \to q^{-1}$ and a rescaling of the 
spectral parameter. 

The other interesting examples of the Q-operators are the ones for the 
conformal filed theory (CFT). The monodromy matrix of the CFT 
can be expressed as an ordered exponential of the form 
$\overline{\mathcal L}= {\mathcal P} \exp \left(\sum_{i=0}^{M+N-1} \int_{0}^{2 \pi} du \ e_{i} \otimes V_{i}(u) \right) $, where $V_{i}(u) $ are 
  q-vertex operators obeying $V_{i}(u)V_{j}(v)=(-1)^{p(i)p(j)}q^{a_{ij}}V_{j}(v)V_{i}(u)$ for $u>v$  and $e_{i}$ are the 
 generators of ${\mathcal B}_{+}$. Thus,  
if we substitute our q-oscillator realizations of ${\mathcal B}_{+}$ through \eqref{evslgl} into the formula 
and taking the supertrace over the Fock space for ${\mathcal B}_{+}$ 
we will obtain Q-operators for the CFT. Examples of such Q-operators 
can be seen for $(M,N)=(2,0)$ in \cite{BLZ97}, 
$(M,N)=(3,0)$ in \cite{BHK02}, 
$N=0$ in \cite{Kojima08}, 
$(M,N)=(2,1)$ in \cite{BT08} and 
for $U_{q}(C(2)^{(2)})$ in \cite{Kulish:2005qc}.  See also a related recent paper \cite{Ridout:2011wx}. 

Finally, we can define the universal master T-operator \cite{AKLTZ11} by 
\begin{align} 
{\mathbb \tau}(x,t)= 
 \sum_{\lambda} S_{\lambda }(t) \Tb_{\lambda}(x),
\end{align}
where $t=(t_{1},t_{2},\dots) $ are time variables in the KP hierarchy and 
$ S_{\lambda }(t)$ is the Schur function labeled by the Young diagram $\lambda $. 
This is a $\tau$-function of the modified KP hierarchy and 
 allows embedding of the quantum integrable system into the soliton theory. 
Basically, all the functional relations among 
T-and Q-operators in the Hirota form can be derived from this (see \cite{AKLTZ11,Kazakov:2010iu} 
for more details).  
\section{Concluding remarks}
In this paper, we have  developed our preliminary discussions on
 L-operators for the Baxter Q-operators for 
 $U_{q}(\widehat{sl}(2|1))$  \cite{talks,BT08} and $U_{q}(\widehat{gl}(3))$  \cite{BT05}, and 
 generalized them to the higher rank case $U_{q}(\widehat{gl}(M|N))$. 
 The contraction of the algebra related to these L-operators was discussed. 
 The model independent universal Q-operators are defined as supertrace of the universal R-matrix. 
This is a step toward our trial \cite{BT08,T09,Tsuboi:2011iz} (also \cite{Kazakov:2010iu,AKLTZ11})  
to construct systematically Q-operators and Wronskian-like expressions 
of T-operators in terms of them. 
The $L$-operators given in this paper can be building blocks of them. 
Our next task \cite{workinprogress} directly related to this paper will be mainly two fold: 
 to generalize our  q-oscillator realization of the 
L-operators for the Q-operators to all the intermediate ones 
 labeled by any $2^{M+N}$ index set $I$ introduced in \cite{T09}, and 
 to generalize these for more general representations on the quantum space. 
 All these will be basically accomplished by evaluating the universal R-matrix 
 in the light of asymptotic representations of 
the quantum affine algebra \cite{HJ11}. 
We find that a 
 fusion method \cite{BFLMS10,Frassek:2011aa} on L-operators for Q-operators developed for rational models is also helpful for this. 
%

A generalization to the elliptic case is perhaps interesting.  
Although whether the contraction of the Sklyanin algebra 
works is not clear at the moment, 
elliptic L-operators may be given by twists
\footnote{This is based on an averaging procedure on the periods with respect to the spectral parameter. 
A similar procedure may also work to lift rational L-operators to trigonometric ones.} of our trigonometric L-operators
 since the elliptic algebras (for both vertex type models and face type models) 
can be obtained by twists on the quantum affine algebras \cite{Fronsdal:1996gc}. 

The other obvious direction of further development will be a generalization to 
 the other quantum affine superalgebras. 
For this, it will be helpful to characterize  
our $L$-operators as sort of Lax operators for the 
generalized Toda system 
\cite{Jimbo:1985ua} 
in terms of the asymptotic algebra \cite{HJ11} 
and investigate the system in the light of the soliton theory \cite{Kazakov:2010iu,AKLTZ11}. 
\section*{Acknowledgments}
The author would like to thank Vladimir Bazhanov for 
collaboration at the early stage of the present work
\cite{talks,BT08,BT05}, 
  and Sergey Khoroshkin for correspondence and allowing him to refer their unpublished results \cite{Bp}, which the author learned from Bazhanov 
through collaboration. 
He also thanks  Vladimir Kazakov and especially Anton Zabrodin 
for a discussion on the generalization of the 
co-derivative for the rational models to the 
trigonometric ones, 
 Takuya Matsumoto and Hiroyuki Yamane for correspondence, 
Tomasz {\L}ukowski for a discussion on a related topic for the rational case and 
Matthias Staudacher for encouraging him to publish the content of 
our talks \cite{talks}.
The work of the author is supported by 
SFB647 ``Space-Time-Matter''.  
A part of this work was done when he was at (or visited) 
Max-Planck-Institut f\"{u}r Gravitationsphysik, 
Albert-Einstein-Institut;  
Ecole Normale Superieure, LPT;  
Osaka City University Advanced Mathematical Institute;  
Okayama Institute for Quantum Physics; 
Department of Theoretical Physics, RSPE, 
Australian National University; and 
Department of Mathematics and Statistics, 
The University of Melbourne
\footnote{He is supported by the Australian
Research Council at ANU and The University of Melbourne.}. 
\\
\\
{\bf Note added for  arXiv:1205.1471v2}
\\
In this version, we made some revisions to the version 1 (arXiv:1205.1471v1) of our paper. 
The revisions are mainly devoted to corrections of misprints 
and additions of details.  
Although the q-oscillator representations of the Borel subalgebra ${\mathcal B}_{+}$ 
for the Q-operators can not be straightforwardly  extended to the whole algebra $U_{q}(\hat{gl}(M|N))$, 
they still can be extended to those of the contracted algebra of $U_{q}(\hat{gl}(M|N))$. 
In version 1, we exemplified this by considering contracted commutation relations \eqref{ef-cont} and  
a part of the intertwining relations (that accompany the co-product and the opposite co-product) for 
the generators $f_{i}$, which come from the other side of 
the Borel subalgebra ${\mathcal B}_{-}$ (after the renormalization), in addition to the generators of ${\mathcal B}_{+}$. 
 In this version, we made these more precise by adding some details. 
 The fact that Serre-type relations for oscillator representations for the Q-operators can be simpler than the original ones 
was pointed out first by \cite{BHK02} for  ${\mathcal B}_{+}$ of $U_{q}(\hat{sl}(3))$ . 
However, a systematic study on this (for $U_{q}(\hat{gl}(M|N;I))$) was missing in the literatures. 
 After version 1 of our paper appeared in May 2012, we received a note from Alessandro Torrielli in November 2012. 
 He discussed an algebra (the co-product, Serre-type relations, etc.) 
related to L-operators for the Q-operators associated with Yangian $Y(sl(2))$. 
However, it is not very clear at the moment how (or if) his result is related to our's. 
\\
{\bf Note added for  arXiv:1205.1471v3}
\\
We improved presentation of the manuscript based on a referee's report. 
We did not add essentially new material to this version. 
Then, the revised manuscript was published at 
 http://dx.doi.org/10.1016/j.nuclphysb.2014.06.017. 
 In addition, we have made the following corrections to the published version:  
 $\lambda_{k} + N-M-k \Rightarrow \lambda_{k} + N+M-k $ in (4.12), 
 $-d_{k}  \Rightarrow -d_{k} -2(N-M+(-1)^{p(k)})$ in (4.19), 
  $x \to zq^{2(M-N)} \Rightarrow x \to xq^{2(M-N)}$ in the footnote 31. 
\\

\section*{Appendix A: Relations for $U_{q}(gl(M|N))$}
\label{relationsglmn}
\addcontentsline{toc}{section}{Appendix A}
\def\theequation{A\arabic{equation}}
\setcounter{equation}{0}
\eqref{rll5} can be rewritten as:
\begin{align}
& [L_{cd}, L_{ab}]=0 \qquad \text{for} \quad 
b<d \le c <a \quad  \text{or} \quad 
d<b \le a < c \quad  \text{or} \quad 
d \le c < b \le a \nonumber \\
& \hspace{115pt} \text{or} \quad 
b \le a < d \le c , 
 \label{relationglmn1}
\\[5pt]
& [L_{cd}, L_{ab}]=(-1)^{(p(a)+p(b))p(c)+p(a)p(b)} (q-q^{-1}) L_{ad} L_{cb} 
\qquad \text{for} \quad 
d <  b  \le c <a ,
\\[5pt]
& [L_{ab}, L_{ad}]_{q^{2p(a)-1}}=0 \qquad \text{for} \quad 
d <b  \le a,
\\[5pt]
& [L_{cb}, L_{ab}]_{q^{1-2p(b)}}=0 \qquad \text{for} \quad 
b \le c < a,
\\[5pt]
& (L_{ab})^2 =0 \qquad \text{for} \quad p(a)+p(b)=1. 
\end{align}
\eqref{rll6} can be rewritten as:
\begin{align}
& [\Lo_{cd}, \Lo_{ab}]=0 \qquad \text{for} \quad 
a<c \le d <b \quad  \text{or} \quad 
c<a \le b < d \quad  \text{or} \quad 
a \le b < c \le d \nonumber \\
& \hspace{115pt} \text{or} \quad 
c \le d < a \le b ,
\\[5pt]
& [\Lo_{ab}, \Lo_{cd}]=(-1)^{(p(a)+p(b))p(d)+p(a)p(b)} (q-q^{-1}) \Lo_{ad} \Lo_{cb} 
\qquad \text{for} \quad 
a <  c  \le b <d , 
\label{relationglmn7}
\\[5pt]
& [\Lo_{ad}, \Lo_{ab}]_{q^{2p(a)-1}}=0 \qquad \text{for} \quad 
a \le b  < d,
\\[5pt]
& [\Lo_{cb}, \Lo_{ab}]_{q^{1-2p(b)}}=0 \qquad \text{for} \quad 
c < a \le b,
\\[5pt]
& (\Lo_{ab})^2 =0 \qquad \text{for} \quad p(a)+p(b)=1. 
\end{align}
\eqref{rll7} can be rewritten as:
\begin{align}
& [L_{cd}, \Lo_{ab}]=0 \qquad \text{for} \quad 
d<a \le b <c \quad  \text{or} \quad 
a<d \le c < b \quad  \text{or} \quad 
d \le c < a \le b \nonumber \\
& \hspace{115pt} \text{or} \quad 
a \le b < d \le c 
\quad \text{or} \quad 
a=b=c=d,
\\[5pt]
& [L_{cd}, \Lo_{ab}]=(-1)^{(p(a)+p(b))p(c)+p(a)p(b)} (q-q^{-1}) \Lo_{ad} L_{cb} 
\qquad \text{for} \quad 
a \le d < b <c \nonumber \\
& \hspace{288pt}   \text{or} \quad a < d < b \le c,
\\[5pt]
& [L_{cd}, \Lo_{ab}]=(-1)^{(p(a)+p(b))p(c)+p(a)p(b)+1} (q-q^{-1}) L_{ad} \Lo_{cb} 
\qquad \text{for} \quad 
d \le a < c <b \nonumber \\
& \hspace{288pt}   \text{or} \quad d < a < c \le b,
\\[5pt]
& [L_{ba}, \Lo_{ab}]=(-1)^{p(b)} (q-q^{-1}) (\Lo_{aa} L_{bb}-L_{aa} \Lo_{bb})
\qquad \text{for} \quad a < b, 
\label{relationglmn14}
\\[5pt]
& [L_{ad}, \Lo_{ab}]_{q^{2p(a)-1}}=0 \qquad \text{for} \quad 
d \le a \le b \quad \text{and} \quad d \ne b,
\\[5pt]
& [L_{cb}, \Lo_{ab}]_{q^{1-2p(b)}}=0 \qquad \text{for} \quad 
a \le b \le c \quad \text{and} \quad a \ne c. 
\label{relationglmn15}
\end{align}
The relations for the contracted algebra $U_{q}(gl(M|N;I))$ can be obtained by applying 
\eqref{red1}-\eqref{red3} for the above relations. 
\section*{Appendix B: Renormalization of generators}
\label{reno-gene}
\addcontentsline{toc}{section}{Appendix B}
\def\theequation{B\arabic{equation}}
\setcounter{equation}{0}
The effect of the renormalization for the L-operator 
\eqref{renoL} to the generators of $U_{q}(\hat{gl}(M|N))$ can be seen from 
\eqref{rll-ev1}, \eqref{evslgl}, \eqref{evglgl} and \eqref{evglgl2}: 
\begin{align}
\tilde{e}_{i}&=e_{i}, 
 \label{e-reno}
\\
\tilde{f}_{i}&=q^{(2-\theta(i \in I)-\theta(i+1 \in I) )m} f_{i}, 
\label{f-reno}
\\
\tilde{h}_{i} &=h_{i}-(\theta(i \in I)-\theta(i+1 \in I))m,
\label{h-reno}
\\
\tilde{k}_{i} &=k_{i}-(-1)^{p(i)}\theta(i \in I)m,
\label{k-reno}
\\
\tilde{\overline{k}}_{i} &=\overline{k}_{i}+(-1)^{p(i)} (2-\theta(i \in I))m,
\label{kbar-reno}
\end{align}
where the right hand side of these should be understood under the evaluation map 
$\mathsf{ev}_{xq^{2m}}$
\eqref{evslgl}-\eqref{evglgl};  
the effect of the renormalization is denoted by tilde; 
and the suffix $i$ should be interpreted under modulo $M+N$. 
\eqref{h-reno} and \eqref{k-reno} came from the transformations for the shift automorphisms  
\eqref{shiftauto} and \eqref{shiftgl}, respectively.  
Then the commutation relations become 
\begin{align}
[\tilde{e}_{i},\tilde{f}_{j}] &=\delta_{ij}
 \frac{q^{2(1-\theta(i+1 \in I))m+\tilde{h}_{i} } -q^{2(1-\theta(i \in I))m-\tilde{h}_{i}} }{q-q^{-1}} . 
  \label{commu-reno2}
  \\
&=\delta_{ij}
 \frac{q^{ (-1)^{p(i)}\tilde{k}_{i} + (-1)^{p(i+1)}\tilde{\overline{k}}_{i+1} } - 
 q^{ (-1)^{p(i)}\tilde{\overline{k}}_{i} + (-1)^{p(i+1)} \tilde{k}_{i+1}} }{q-q^{-1}} . 
  \label{commu-reno3}
\end{align}
Let us consider the limit $m \to \infty$ for $|q| <1$ (or $m \to -\infty$ for $|q| >1$). 
We assume the renormalized generators except for \eqref{kbar-reno} do not diverge in this limit 
at least for the evaluation representation $\pi_{I}(xq^{2m})$ in an appropriate basis, where $\pi_{I}$ is the highest 
weight representation of $U_{q}(gl(M|N))$ with  
 the highest weight \eqref{highestweight}.  
 Then, in the limit, we obtain:
\begin{align}
q^{ (-1)^{p(i)}\tilde{ k }_{i} } q^{ (-1)^{p(i)}\tilde{ \overline{k} }_{i} } = q^{2m(1-\theta(i \in I))} \to  \theta (i \in I ),  
\end{align}
and  in particular
\begin{align}
q^{ (-1)^{p(i)}\tilde{ \overline{k} }_{i} } \to 0 \qquad \text{for} \quad i \in \overline{I} . 
\end{align}
The inverse of $q^{ (-1)^{p(i)}\tilde{ k }_{i} }$, namely $q^{- (-1)^{p(i)}\tilde{ k }_{i} }$ coincides with $q^{ (-1)^{p(i)}\tilde{ \overline{k} }_{i} } $ 
only for $i \in I$ in the limit. 
Then the commutation relations \eqref{commu-reno2} 
reduce to the contracted commutation relations \eqref{ef-cont} in the limit. 
Note that the limit of 
\eqref{commu-reno2} automatically hold true if $\tilde{f}_{i}=0 $ for $i, i+1 \in \overline{I}$ 
in the limit. 

Let us multiply ($U_{q}(\hat{gl}(M|N))$ case of) the  first relation  in \eqref{R-def}   for $f_{i}$
by $q^{(2-\theta(i \in I)-\theta(i+1 \in I) )m}(1 \otimes q^{- m \sum_{j \in I} k_{j}}) $ from the right: 
\begin{multline}
 \left(
 q^{2(1-\theta(i+1 \in I) )m}
q^{\tilde{h}_{i}  }  \otimes f_{i} 
+ \tilde{f}_{i}\otimes 1 \right) \tilde{\mathcal R} ( 1 \otimes q^{-m \sum_{j \in I} k_{j}} )
= \\
=
\tilde{\mathcal R}  (1 \otimes q^{-m \sum_{j \in I} k_{j}} )
\left( 
 \tilde{f}_{i}  \otimes q^{h_{i}}  +
  q^{2(1-\theta(i \in I) )m}
 1 \otimes f_{i}  
\right) ,
\end{multline} 
where $\tilde{R}$ is defined in \eqref{R-red-gl}. 
One can see an effect of the shift automorphism for $\mathcal{B}_{+}$ by the transformation 
\eqref{shiftgl} with the parameters \eqref{shiftpara}. 
Then this relation for 
$\pi_{I}(xq^{2m}) \otimes \pi(y)  $ suggests 
 \eqref{intert3} in the limit. 


\begin{thebibliography}{99}
\bibitem{BT08}
V.V.\ Bazhanov, Z.\ Tsuboi: 
Baxter's Q-operators for supersymmetric spin chains, 
Nucl. Phys. B 805 [FS] (2008) 451-516 [arXiv:0805.4274 [hep-th]]
\footnote{ 
We were pointed out by Carlo Meneghelli  that our L-operators pose a problem in the rational limit. 
We found that this is due to the following misprint. 
In  page 465, eq. (2.86), 
$C_{f}=q^{-1}q^{-2{\mathcal H}_{f}}(1-f^{+}f^{-})$ is a misprint of 
$C_{f}=q q^{-2{\mathcal H}_{f}}(f^{+}f^{-}+q^{-2}f^{-}f^{+})
= q^{-1}q^{-2{\mathcal H}_{f}}(1+(q^2-1)f^{+}f^{-})$. 
We thank him for his kind comment.
}. 
%

\bibitem{T09}
Z. Tsuboi, 
Solutions of the $T$-system and Baxter equations for supersymmetric spin chains, 
Nucl. Phys. B 826 [PM] (2010) 399-455  [arXiv:0906.2039 [math-ph]]. 

\bibitem{Tsuboi:2011iz} 
  Z.~Tsuboi,
  ``Wronskian solutions of the T, Q and Y-systems related to infinite dimensional unitarizable modules of the general linear superalgebra $gl(M|N)$,'' Nucl.\ Phy.\ B 870 [FS] (2013) 92-137 [arXiv:1109.5524 [hep-th]].

\bibitem{Bax72}
R.J.\ Baxter, 
Partition function of the eight-vertex lattice model, 
 Ann. Phys. 70 (1972) 193-228.
 
\bibitem{BLZ97}
%
V.V.\ Bazhanov, S.L.\ Lukyanov, A.B.\ Zamolodchikov, 
Integrable Structure of Conformal Field Theory III. The Yang-Baxter Relation, 
Commun.Math.Phys. 200 (1999) 297-324 
[arXiv:hep-th/9805008]. 

\bibitem{BHK02}
V.V.\ Bazhanov, A.N.\ Hibberd, S.M.\ Khoroshkin, 
 Integrable structure of ${W}_3$ Conformal Field Theory, Quantum
  {B}oussinesq Theory and Boundary Affine {T}oda Theory, 
Nucl. Phys. B622 (2002) 475--547 [arXiv:hep-th/0105177].

\bibitem{Kulish:2005qc}
  P.~P.~Kulish, A.~M.~Zeitlin,
  ``Superconformal field theory and SUSY N=1 KDV hierarchy II: The Q-operator,''
  Nucl.\ Phys.\  {\bf B709 } (2005)  578 
  [hep-th/0501019].
 
\bibitem{Boos07}
H.\ Boos, M.\ Jimbo, T.\ Miwa, F.\ Smirnov, Y.\ Takeyama, 
 Hidden Grassmann structure in the XXZ model, 
Commun. Math. Phys. 272 (2007) 263-281 [arXiv:hep-th/0606280]. 

\bibitem{Kojima08}
T.\ Kojima,
\textit{``The Baxter's Q-operator for the W-algebra \(W_N\),''}
J.Phys.A: Math. Theor. 41 (2008) 355206
[arXiv:0803.3505 [nlin.SI]].

\bibitem{BGKNR10}
H.\ Boos, F.\ G\"{o}hmann, A.\ Kl\"{u}mper, K.S.\ Nirov, A.V.\ Razumov, 
Exercises with the universal R-matrix 
J. Phys. A: Math. Theor. 43 (2010) 415208 [arXiv:1004.5342 [math-ph]].
%

\bibitem{GKLT10}
N.\ Gromov, V.\ Kazakov, S.\ Leurent, Z.\ Tsuboi, 
Wronskian Solution for AdS/CFT Y-system, 
JHEP 1101 (2011) 155 [arXiv:1010.2720 [hep-th]]. 

%
\bibitem{Faddeev:1987ih} 
  L.~D.~Faddeev, N.~Y.~.Reshetikhin and L.~A.~Takhtajan,
  ``Quantization of Lie Groups and Lie Algebras,''  Leningrad Math.\ J.\  {\bf 1}, 193-225 (1990) 
 [Alg.\ Anal.\  {\bf 1}, 178-206
 (1989)].
 
\bibitem{Bp}
V.V.\ Bazhanov, private communication (2005): 
V.V.\ Bazhanov, S.M.\ Khoroshkin, (2001) unpulished.

\bibitem{talks}
V.V.\ Bazhanov, Z.\ Tsuboi: 
%
%
talks at conferences in 2007, which include the following two: 
%
La 79eme Rencontre entre physiciens theoriciens et mathematiciens 
``Supersymmetry and Integrability'', IRMA Strasbourg, June, 2007  [http://www-irma.u-strasbg.fr/article383.html]; 
%
``Workshop and Summer School: 
From Statistical Mechanics to Conformal and Quantum Field Theory'', 
the university of Melbourne, January, 2007 
[http://www.smft2007.ms.unimelb.edu.au/program/LectureSeries.html]. 
%

\bibitem{HJ11}
D.\ Hernandez, M.\ Jimbo,
Asymptotic representations and Drinfeld rational fractions, 
Compos.\ Math.\ 148 (2012) 1593-1623
[arXiv:1104.1891 [math.QA]].

\bibitem{BT05}
V.V.\ Bazhanov, Z.\ Tsuboi: unpublished (2005).


\bibitem{BFLMS10}
V. Bazhanov, R. Frassek, T. Lukowski, C. Meneghelli, M. Staudacher, 
Baxter Q-Operators and Representations of Yangians, 
Nucl.Phys. B850 (2011) 148-174 [arXiv:1010.3699 [math-ph]]; 
%
\\
R. Frassek, T. Lukowski, C. Meneghelli, M. Staudacher, 
Oscillator Construction of $su(n|m)$ Q-Operators, 
Nucl. Phys. B 850 (2011) 175-198 
[arXiv:1012.6021 [math-ph]]. 

\bibitem{BDKM06}
A.V.\ Belitsky, S.E.\ Derkachov, G.P.\ Korchemsky, A.N.\ Manashov:
 Baxter Q-operator for graded $SL(2|1)$ spin chain,
 J.Stat.Mech. 0701 (2007) P005 [arXiv:hep-th/0610332]; 
%
\\
S.\ E.\ Derkachov, A.\ N.\ Manashov, 
Factorization of R-matrix and Baxter Q-operators for generic $sl(N)$ spin chains,
J.Phys.A42 (2009) 075204 [arXiv:0809.2050 [nlin.SI]]. 

\bibitem{Frassek:2011aa} 
  R.~Frassek, T.~Lukowski, C.~Meneghelli and M.~Staudacher,
  ``Baxter Operators and Hamiltonians for 'nearly all' Integrable Closed $gl(n)$ Spin Chains,'' 
Nucl.\ Phys.\ B 874 [PM] (2013) 620-646 
 [arXiv:1112.3600 [math-ph]].
 
\bibitem{Kazakov:2010iu} 
  V.~Kazakov, S.~Leurent and Z.~Tsuboi,
  ``Baxter's Q-operators and operatorial Backlund flow for quantum (super)-spin chains,''  Commun.\ Math.\ Phys.\ 311(2012) 
787-814  [arXiv:1010.4022 [math-ph]].

\bibitem{KV07}
  V.~Kazakov and P.~Vieira,
  ``From Characters to Quantum (Super)Spin Chains via Fusion,''
  JHEP 0810 (2008) 050
  [arXiv:0711.2470 [hep-th]].
  
\bibitem{AKLTZ11}
A.\ Alexandrov, V.\ Kazakov, S.\ Leurent, Z.\ Tsuboi, A.\ Zabrodin, 
Classical tau-function for quantum spin chains, 
JHEP 1309 (2013) 064
[arXiv:1112.3310[math-ph]]. 
\bibitem{Yamane99}
 H. Yamane, On defining relations of affine Lie superalgebras and
 affine quantized universal enveloping superalgebras,
 Publ. Res. Inst. Math. Sci. 35 (1999) 321-390;  errata:
 Publ. Res. Inst. Math. Sci. 37 (2001) 615--619 [arXiv:q-alg/9603015]; 
H. Yamane,
Examples of the defining relations of the quantum
affine superalgebras, 
http://www16.tok2.com/home/hiroyukipersonal/pdf1.pdf  

\bibitem{KT94}
S. Khoroshkin, V. Tolstoy,  Twisting of quantum
(super)algebras. Connection of Drinfeld's and Cartan-Weyl realizations
for quantum affine algebras [arXiv:hep-th/9404036]. 

\bibitem {Dr85} 
V. Drinfeld, Hopf algebras and the quantum Yang-Baxter equations,
Sov. Math. Dokl. 32 (1985) 264-268.

\bibitem{KT92}
S. M. Khoroshkin and  V. N. Tolstoy, The uniqueness theorem for the
universal ${R}$-matrix, Lett. Math. Phys. 24 (1992) 231--244. 

\bibitem{KT91}
S. M. Khoroshkin and  V. N. Tolstoy,
Universal R-matrix for quantized (super)algebras, 
Commun.\ Math.\ Phys.\ 141 (1991) 599-617.

\bibitem{Perk:1981nb} 
  J.~H.~H.~Perk and C.~L.~Schultz,
  ``New families of commuting transfer matrices in q state vertex models,''  
Phys.\ Lett.\ A {\bf 84}, 407-410 (1981). 


\bibitem{Cherednik80}
I. V. Cherednik, On a method of constructing factorized $S$ matrices in
        elementary functions, Theor. Math. Phys. 43 (1980) 356-358.    
 
\bibitem{FM01}
E.\ Frenkel, E.\ Mukhin, 
The Hopf algebra $Rep U_q \hat{gl}_\infty$, 
Selecta Mathematica, New Series 8 (2002) 537-635 
[arXiv:math/0103126 [math.QA]]; 
\\ 
M. J. Hopkins, A. I. Molev, 
A q-Analogue of the Centralizer Construction and Skew Representations of the Quantum Affine Algebra, 
SIGMA 2 (2006) 092 [arXiv:math/0606121 [math.QA]].   

%
\bibitem{Chaichian:1989rq} 
  M.~Chaichian and P.~Kulish,
  ``Quantum Lie Superalgebras and q-Oscillators,''  Phys.\ Lett.\ B {\bf 234}, 72-80 (1990). 

\bibitem{KNS93}
A.\ Kuniba, T.\ Nakanishi, J.\ Suzuki, 
Functional Relations in Solvable Lattice Models I: Functional
Relations and Representation Theory, 
Int. J. Mod. Phys. A9 (1994) 5215-5266 [arXiv:hep-th/9309137]. 


\bibitem{T97}
Z.\ Tsuboi, Analytic Bethe Ansatz and functional equations 
 for Lie superalgebra $sl(r+1|s+1)$, 
J.Phys.A: Math. Gen. 30 (1997) 7975-7991[arXiv:0911.5386 [math-ph]]; 
Z.\ Tsuboi, 
Analytic Bethe Ansatz and functional equations associated with 
 any simple root systems of the Lie superalgebra 
 $sl(r+1|s+1)$, Physica A 252 (1998) 565-585 [arXiv:0911.5387 [math-ph]]. 

\bibitem{Ridout:2011wx} 
  D.~Ridout and J.~Teschner,
  ``Integrability of a family of quantum field theories related to sigma models,''  Nucl.\ Phys.\ B {\bf 853}, 
327-378 (2011)  [arXiv:1102.5716 [hep-th]].

\bibitem{KLWZ97}
I.\ Krichever, O.\ Lipan, P.\ Wiegmann, A.\ Zabrodin, 
 Quantum Integrable Models and Discrete Classical Hirota Equations, 
Commun. Math.Phys. 188 (1997) 267-304 [arXiv:hep-th/9604080].

\bibitem{Kazakov:2007fy} 
  V.~Kazakov, A.~S.~Sorin and A.~Zabrodin,
  ``Supersymmetric Bethe ansatz and Baxter equations from discrete Hirota dynamics,''  Nucl.\ Phys.\ B {\bf 790}, 345-413 (2008)  [hep-th/0703147 [HEP-TH]].

%

\bibitem{workinprogress}
work in progress.
  
\bibitem{Fronsdal:1996gc} 
  C.~Fronsdal,
  ``Quasi-Hopf deformations of quantum groups,''  Lett.\ Math.\ Phys.\  {\bf 40}, 117-134 (1997)  [arXiv:q-alg/9611028]; 
%
C.~Fronsdal, 
Generalization and Exact Deformations
of Quantum Groups, Publ. RIMS, Kyoto Univ.
33 (1997), 91-149 [arXiv:q-alg/9606020]. 

\bibitem{Jimbo:1985ua} 
  M.~Jimbo,
  ``Quantum $R$ Matrix for the Generalized Toda System,''  Commun.\ Math.\ Phys.\  {\bf 102}, 537-547 (1986);
\\
  V.~V.~Bazhanov,
  ``Integrable Quantum Systems and Classical Lie Algebras,
''  Commun.\ Math.\ Phys.\  {\bf 113}, 471-503 (1987).




 



\end{thebibliography}
\end{document}